# A Hybrid Post-Quantum Encryption Architecture with Self-Hosted Key Management for SME Cloud Data Protection

Muhammad Shaheer Bin Junaid
Independent Researcher
Karachi, Pakistan
*contact@mshaheerbjunaid.com*
ORCID: 0009-0002-4088-6232

**Abstract—** Harvesting ciphertext from cloud storage needs no quantum computer; decrypting it later does. That gap is the harvest-now-decrypt-later exposure: anything protected by RSA or ECDH today that must stay secret for decades is already compromised. Small and medium-sized enterprises are least able to respond: they neither run the infrastructure on which their data sits on nor employ a cryptographer. Bespoke migration suits firms with security budgets; a managed key service relocates trust rather than removing it. The obstacle is architectural, not cryptographic. We present *Quantum Cloud Guard* (QCG), a software-only three-layer architecture. No prior SME-oriented system combines its three elements: client-side hybrid post-quantum encryption, self-hosted key custody with client-verifiable ML-DSA-87 signatures on served keys, and an integrated application-layer abuse-prevention gateway. Files never leave the client: each is sealed under AES-256-GCM, its key wrapped to an ML-KEM-1024 public key from the enterprise's key service. The enterprise alone administers it; it signs every key with ML-DSA-87, so a client that pinned it detects substitution. Separating *key custody from data custody* is the point: a provider holding both can read the data. On a 24 MHz STM32F407, ML-KEM-1024 key generation takes 40.8 ms and decapsulation 44.0 ms; on the server every post-quantum operation stays sub-millisecond, signing adding 0.24 ms per request. The service runs on a 4.49 EUR/month virtual server. Under sustained flooding, the in-process gateway Sentinel Gate rejected 98.8% of attack traffic while a legitimate client's median latency moved from 621 to 625 ms. Being single-source, this shows filtering effectiveness, not DDoS resilience.

## I. Introduction

More than half the world's data now sits in the cloud [1]. For small and medium-sized enterprises (SMEs), dependence on Amazon Web Services, Google Cloud Platform, or Microsoft Azure is not a choice so much as a cost structure, and that dependence has opened an exposure current encryption practice does not close.

The public-key cryptography underneath, RSA and elliptic curves, rests on problems classical computers cannot solve efficiently [2], and for decades that was enough. Shor ended the assumption in 1994: a large enough quantum computer breaks both [3]. The threat used to look remote. A 2025 result cut the estimated qubit count for breaking RSA-2048 from roughly twenty million to under one million [4], and the most recent expert survey rates a cryptographically relevant quantum computer as quite possible within ten years and likely within fifteen [5].

Waiting for that machine is not the risk. The attack has already started: under harvest-now-decrypt-later (HNDL), an adversary records encrypted data today and decrypts it once the hardware arrives [6], [17], so anything with a long confidentiality life (health records, financial data, legal files, proprietary designs) is being collected now. NIST names HNDL directly in its transition guidance [7]. After an eight-year standardisation effort, it published its first three post-quantum standards in August 2024: FIPS 203 (ML-KEM, from *CRYSTALS-Kyber*), FIPS 204 (ML-DSA, from *CRYSTALS-Dilithium*), and FIPS 205 (SLH-DSA, from SPHINCS+) [8], with FN-DSA (from FALCON) to follow as the draft FIPS 206.

The government has since moved from guidance to mandate. Executive Order 14412, signed in June 2026, requires federal high-value assets and high-impact systems to migrate to post-quantum key establishment by the end of 2030 and to post-quantum authentication by the end of 2031, and extends the same obligation to federal contractors [63]. The separation of those two deadlines is instructive: authentication is the harder migration and the one least advanced in practice. Industry is moving on the same clock, Google targeting full post-quantum migration by 2029 [9] and AWS building quantum-safe services to NIST guidance [10]. The SME sector is not. It makes up more than ninety per cent of businesses worldwide. It employs almost sixty per cent of the workforce [11], yet fewer than five per cent of enterprises have a formal quantum-transition plan [13]. The usual constraints apply: thin security budgets, no cryptographer on staff, dependence on third-party clouds, and no framework written for firms this size [12]. Managed post-quantum key services look like the answer, but they are not. The threat model treats the cloud provider as an untrusted host of long-lived data; handing that same provider the keys reintroduces the trust the architecture exists to remove, and does so under per-operation pricing that a self-hosted service avoids.

The providers secure their own estates. What they do not secure is the encryption an SME performs before its data leaves the building, and that step is almost always classical. The literature offers hybrid classical and post-quantum frameworks [14], hybrid TLS [15], and quantum key distribution paired with AES-256 [16], but these target enterprise systems or need dedicated hardware such as a quantum channel, and the more theoretical proposals give a hardware-limited SME no software-only path it can actually deploy.

This paper introduces Quantum Cloud Guard (QCG), a hybrid three-layer encryption architecture for SMEs on public or third-party clouds. QCG needs no specialised quantum hardware, runs over the cloud infrastructure the firm already uses, and leaves key control with the SME rather than the provider. The first layer is client-side hybrid encryption: each file is sealed with AES-256-GCM, and the AES-256 key is encapsulated under ML-KEM-1024 (formerly CRYSTALS-Kyber), a NIST-standardised key-encapsulation mechanism (KEM). The split is deliberate. A symmetric cipher that holds against quantum search protects the data; a post-quantum mechanism protects the key, which is the piece an HNDL adversary actually wants; and an attacker must defeat both, so the design survives even if flaws surface in the newer lattice schemes. The second layer is a Quantum-Safe Key Management Service (QS-KMS) on an inexpensive virtual private server (VPS). It generates ML-KEM-1024 key pairs, stores secret keys encrypted, and issues short-lived session tokens that let a user authenticate from any device without a key ever reaching the cloud; tokens are revoked server-side when the client closes, with a hard one-hour ceiling. Sentinel Gate, a Layer 7 abuse-prevention gateway built into the QS-KMS, adds per-identity rate limiting, dynamic blacklisting, progressive tarpitting of brute-force attempts, and logging of every access event [39]. ML-DSA-87 (formerly CRYSTALS-Dilithium) supplies integrity and authenticity.

This paper makes three contributions, and none of them is a new cryptographic primitive; the primitives are standard and used as standardised. What is new is the arrangement, and the case that it holds up.

**1) The QCG architecture.** A software-only, client-side hybrid design that keeps full key custody at the SME and asks for no special hardware. Its novelty is not any primitive but the fit to the budget, expertise, and trust constraints a small firm actually works under, and no prior SME-oriented system combines client-side hybrid post-quantum encryption, self-hosted key custody with client-verifiable ML-DSA-87 signatures on every served key, and an integrated application-layer abuse-prevention gateway.

**2) A self-hosted key service defended as a first-class concern.** A QS-KMS whose availability and integrity are protected by the integrated Layer 7 gateway. Prior SME-focused post-quantum frameworks secure confidentiality and stop there, leaving their own key services exposed. QCG treats protection of the key infrastructure as a security property in its own right, through per-identity rate limiting, resistance to credential stuffing and brute force, and single-source flood mitigation.

**3) An empirical account of the cost.** What quantum-safe protection actually costs under SME-realistic conditions: a performance floor for ML-KEM-1024 on constrained ARM Cortex-M4 hardware, and end-to-end measurements across file sizes on a live deployment, which together show post-quantum data protection to be computationally and economically feasible at SME scale.

Hybrid post-quantum encryption already exists in several forms. What does not exist is a single architecture that satisfies, at once, the five constraints a small enterprise actually works under:

**(i) Software alone,** with no specialised hardware;

**(ii) Key custody held apart from data custody,** so the party storing the ciphertext cannot read it;

**(iii) A key service defended at the application layer,** not left exposed once confidentiality is handled;

**(iv) A flat, predictable cost** that does not penalise good key hygiene;

**(v) A working, inspectable implementation,** not a design on paper.

The prior work surveyed in the next section satisfies at most two of these together. QCG is built for all five, and the conjunction, not any single primitive, is the contribution.

The rest of the paper runs as follows. Section II describes the changing quantum threat. Section III reviews related work. Section IV sets out the methodology. Section V covers the cryptographic background. Sections VI and VII describe the framework layers. Section VIII presents the proof-of-concept implementation; Section IX the evaluation and benchmarks; Section X the implementation and deployment challenges; and Section XI the security analysis. Section XII offers a broader discussion and future work, and Section XIII concludes.

## II. Evolving Threats & Quantum Computing

Every cipher in production rests on a bet: that certain mathematical problems stay out of reach of any machine an adversary can build. Quantum computers [18] call that bet for one class of problems and leave it standing for another. The distinction matters because it decides what must be replaced and what merely reinforced. It does not exhaust the danger either: several of the attacks a deployed design has to survive need no quantum computer at all, and this section covers those too.

### A. Shor's Algorithm and Public-Key Cryptography

RSA, elliptic-curve cryptography, and Diffie-Hellman all lean on two problems, integer factorisation and the discrete logarithm, that classical computers can attack only in sub-exponential time. Shor showed in 1994 that a quantum computer solves both in polynomial time [3]. Nothing about that result is gradual. Once a large enough machine exists, all three schemes fail together, and with them go the mechanisms secure communication uses to prove identities and move keys across networks. There is no parameter to tune and no key length that helps; the problem itself has stopped being hard.

### B. Grover's Algorithm and Symmetric Security

Grover's algorithm threatens the symmetric side far less. Searching an unstructured space, it gains only a quadratic speed-up [19], so brute force against an $n$-bit key drops from order $2^n$ to order $2^{n/2}$. Double the key length and the advantage is gone. AES-256 keeps an effective 128 bits against quantum search, comfortably beyond practical reach; AES-128 falls to an effective 64 bits, which is not. Some analyses argue that the cost of actually running Grover at scale makes even the halving pessimistic [20]. Either way, a 256-bit key leaves margin to spare. This asymmetry, catastrophic for public-key schemes and manageable for symmetric ones, is the foundation the present design stands on: replace the broken public-key component, keep a symmetric cipher of adequate length.

### C. Quantum Impact Across Cryptographic Primitives

Not every primitive meets the same fate, then. Some break outright, some lose strength, and some need nothing more than a longer key. Table I sorts the common primitives into these categories and states the response each one demands.

**TABLE I. IMPACT OF QUANTUM ALGORITHMS ON COMMON CRYPTOGRAPHIC PRIMITIVES**

| Cryptographic Primitive | Type | Relevant Quantum Algorithm | Effect | Required Response |
|---|---|---|---|---|
| RSA | Public-key | Shor’s | Broken | Replace with PQC |
| ECC / ECDSA | Public-key | Shor’s | Broken | Replace with PQC |
| Diffie-Hellman | Key exchange | Shor’s | Broken | Replace with PQC KEM |
| AES-128 | Symmetric | Grover’s | Weakened (approx. 64-bit) | Increase key size |
| AES-256 | Symmetric | Grover’s | Mildly weakened (approx. 128-bit) | Considered safe |
| SHA-256 | Hash | Grover’s | Mildly weakened (approx. 128-bit) | Considered safe |
| SHA-3 | Hash | Grover’s | Mildly weakened | Considered safe |

One pattern stands out. Everything in the broken rows is machinery for establishing trust and for exchanging or wrapping keys; everything that survives is doing bulk encryption. A quantum-safe system therefore has to be rebuilt exactly where keys are agreed and identities proven, and it can keep its symmetric core. That is the shape QCG takes, and Fig. 1 map each primitive to its post-quantum replacement.

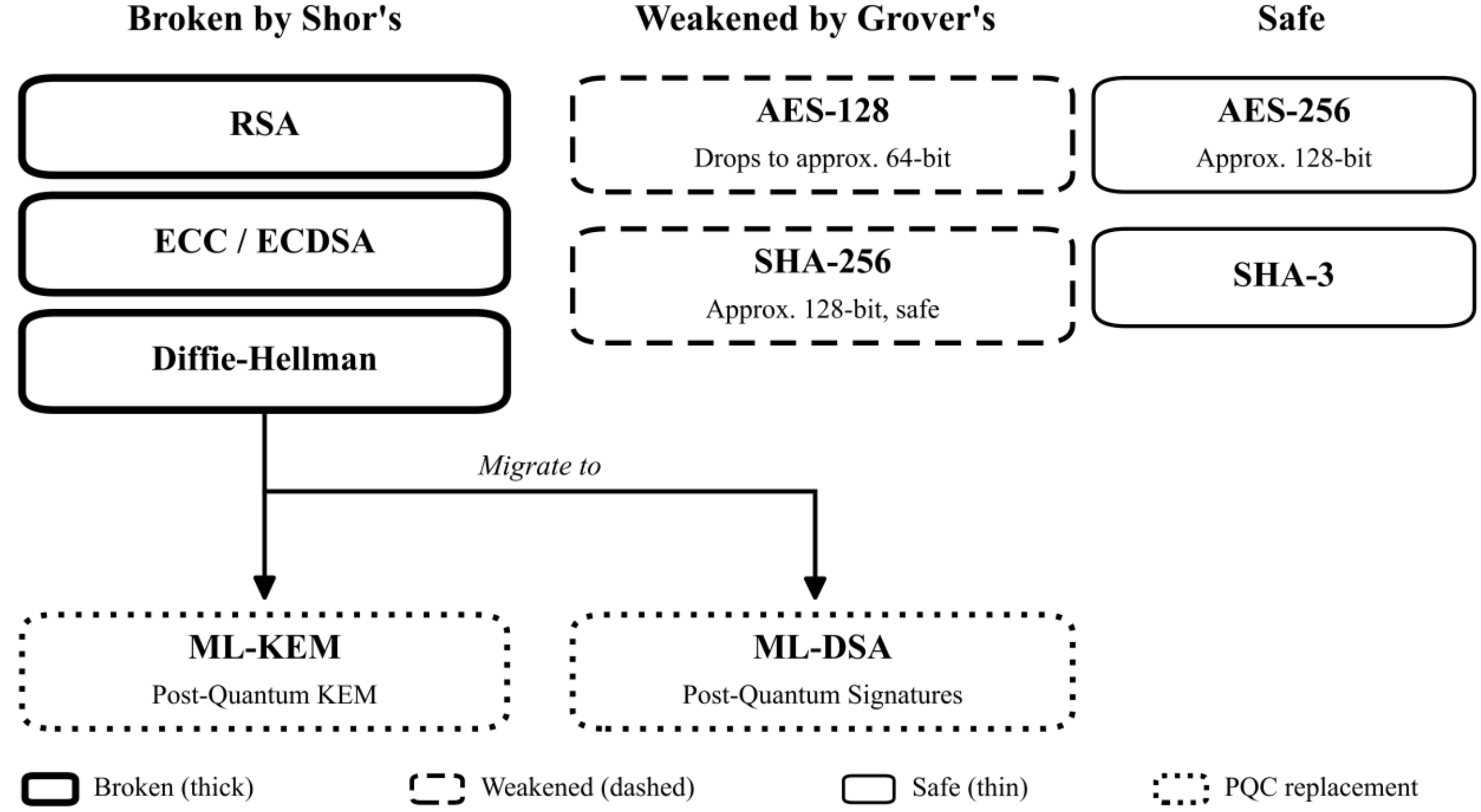

*Fig. 1. Quantum threat landscape across cryptographic primitives and their post-quantum migration paths.*

### D. The Broader Threat Landscape

It would be a mistake to compress all of this into the sentence that quantum computers will break encryption, because a deployable framework has to survive attacks that do not wait for one. The most pressing is harvest-now-decrypt-later: record the ciphertext today, decrypt it once the machine exists [17]. Collection needs no quantum capability whatsoever, so any adversary can run the first half of the attack right now, and the data that must stay confidential longest (medical histories, legal files, financial archives) is exactly the data most exposed. The moment of key establishment carries its own risk, since an adversary who can solve the underlying problem can recover the session key outright or interpose quietly between the two parties. Implementations open a further front. A scheme can be quantum-safe on paper and still bleed key material through timing, power draw, or electromagnetic emission, and the lattice schemes ML-KEM and ML-DSA are known to be sensitive to timing analysis in particular [21]; that weakness lives in the code rather than the mathematics. Finally, a system that offers both a classical and a post-quantum mode can be pushed into negotiating the weaker one. Downgrade attacks of this kind threaten only hybrid, negotiation-based designs, but for them the threat is real.

### E. Why SMEs Are Uniquely Exposed

None of this lands evenly. A large enterprise or a cloud provider has security staff, cryptographic expertise, and a migration budget; a small or medium-sized enterprise usually has none of the three. It runs on cloud services it cannot instruct to migrate, employs nobody who can weigh the threat properly, and cannot pay for a bespoke fix. So a very large population of organisations sits exposed, and every month of delay raises the eventual cost, because harvest-now-decrypt-later collection is presumably already under way. That is the gap the rest of this paper works on.

## III. Related Work

Combining classical and post-quantum mechanisms so that the whole stands while either part stands is, by now, the accepted way through the transition, and the literature on such hybrids is substantial. QCG borrows its techniques from that literature; what it claims as its own is the arrangement. Three streams matter here: hybrid post-quantum encryption schemes, hybrid encryption built for the cloud, and studies of how enterprises migrate. Each supplies a piece, and the gap sits where they fail to meet.

### A. Hybrid Post-Quantum Encryption Schemes

Asif's authenticated post-quantum session protocol [22] is the nearest recent relative. It joins ML-KEM-1024 for key exchange, ML-DSA for authentication, and AES-256-GCM for the data itself, and reports sub-millisecond cryptographic overhead across 1,000 iterations on commodity hardware. At its core sits the same KEM-DEM construction QCG uses in Layer 1, which is reassuring: the approach holds up in independent hands. Where the two works part is the problem they solve. Asif establishes authenticated, short-lived sessions between endpoints that are talking to each other. QCG protects data at rest before it ever reaches a provider it does not trust, and everything that surrounds the shared core (the self-hosted key custody, the encrypted vault, the gateway in front of the key service) exists because data at rest demands what a session protocol never needs.

### B. Hybrid Encryption for Cloud Environments

A second line brings hybrid encryption to cloud data specifically. SymECCipher is a recent case: ECC for the key exchange, AES for the data, arranged client-side for healthcare records in the cloud [23]. The demand it answers is real, and it answers it in the right place, at the client. The trouble is the key exchange, because elliptic-curve cryptography is precisely what Shor's algorithm removes, so the scheme protects against every adversary except the one this paper is about. QCG keeps the client-side hybrid philosophy and swaps the vulnerable exchange for a post-quantum KEM. Beyond that, the cloud stream runs to hybrid TLS and to quantum key distribution paired with symmetric ciphers [15], [16]. Both are serious proposals, and both suit organisations that can afford them: QKD wants dedicated quantum channels, hybrid TLS wants engineering attention, and an ordinary SME, short on budget and shorter on cryptographers, was never their audience.

### C. SME and Enterprise Migration Studies

The third stream studies the organisation rather than the mechanism. One careful recent analysis puts migration at roughly five to seven years for a small business, eight to twelve for a medium one, and longer again above that [24], with budgets, personnel, and vendor timelines as the binding constraints while harvest-now-decrypt-later runs in the background; further surveys find smaller firms broadly unprepared [11], [13]. All of this sharpens the motivation for the present paper. What it does not do is tell any single organisation what to deploy on Monday. It maps where the field must go and how hard the road is; the question of how one SME actually proceeds is the one this paper takes up.

### D. Positioning of QCG

Set side by side, the three streams outline a gap rather than a crowded field, and the shape of the gap is specific. The hybrid constructions are technically sound but built for general secure communication between endpoints. The cloud-oriented schemes move encryption towards the client yet keep a quantum-vulnerable key exchange or lean on specialised hardware. The migration studies name the SME problem with precision [53] and hand over no mechanism for solving it. QCG stands where the three would intersect: an established hybrid KEM and symmetric construction, commodity hardware only, keys entirely in the SME's hands, implemented at the client. The closest implementation on record is Gandhi et al. [47], which applies the same hybrid of ML-KEM and AES-256-GCM to end-to-end messaging through a zero-trust relay. This server forwards ciphertext between clients without ever holding keys or plaintext. It is an elegant design for its purpose, and its purpose is different. The system lives at the application layer, serves real-time communication, and, by its authors' own statement, puts implementation ahead of novelty, deferring authentication and digital signatures to future work and providing no key-management infrastructure at all. QCG differs in domain and in scope. It targets stored cloud data rather than messages in flight, runs the higher ML-KEM-1024 parameter set, and adds precisely what the messaging system leaves out: the self-hosted QS-KMS, ML-DSA authentication of every served public key, and a dedicated access-protection layer in front of the key service. None of these appear together in that earlier work, and among the surveyed SME-oriented systems no other treats the availability of the key infrastructure as a goal equal in standing to confidentiality [39]. Table II summarises the comparison.

**TABLE II. COMPARISON OF QCG WITH REPRESENTATIVE PRIOR WORK**

| Work | Post-Quantum | Cloud Focused | SME Focused | Software-Only (No Special Hardware) | Client-Controlled Keys | Self-Hosted Key Management |
|---|---|---|---|---|---|---|
| Hybrid PQ session protocol [22] | Yes | No | No | Yes | No | No |
| Hybrid PQ end-to-end encryption [47] | Yes | No | No | Yes | Yes | No |
| SymECCipher (ECC-AES) [23] | No | Yes | No | Yes | Partial | No |
| Hybrid TLS / QKD schemes [15] [16] | Yes | Partial | No | No | No | No |
| Enterprise migration analysis [24] | N/A (strategy) | Partial | Partial | N/A | N/A | N/A |
| QCG (this work) | Yes | Yes | Yes | Yes | Yes | Yes |

Read the table column by column and, to the best of our knowledge within the surveyed literature, no single prior system combines the properties QCG targets. Every attribute in Table II exists somewhere; the conjunction, in the SME client-side setting, exists nowhere, and that conjunction, not any new primitive, is what this paper contributes. The hybrid protocols reach post-quantum confidentiality but assume live sessions and bring no key management. The cloud schemes either keep a breakable exchange or ask for hardware an SME will not buy. The migration studies stop before the artefact. The sharpest line of separation is self-hosted custody, because it is exactly what the newly available managed post-quantum key services decline to offer: hand the keys to the provider that stores the data, and the separation the whole threat model rests on is gone.

## IV. Research Methodology

The work follows Design Science Research: build an artefact against a real problem, evaluate it, and treat what the building itself teaches as part of the result. The problem is the one the previous two sections established: SMEs exposed to harvest-now-decrypt-later and poorly served by what exists; the artefact is QCG. The methodology deserves explicit statement because several of the decisions that now distinguish the system were not chosen in advance at all. They were forced, mid-construction, by things that refused to work as designed, and recording those turns is not housekeeping; in this paradigm it is required output. Three elements follow: an incremental construction discipline, the design revisions reality imposed, and an evaluation strategy that spans a constrained microcontroller and a live deployment.

### A. Incremental Construction and Layer-by-Layer Validation

Construction ran strictly bottom-up, and no layer was added until the one beneath it had been proven alone. The order was deliberate. First the cryptographic core, ML-KEM-1024 encapsulation wrapped around AES-256-GCM in the KEM-DEM construction; then the encrypted storage that seals key material at rest; then the service layer that binds storage and cryptography into coherent operations; then the network interface; and last the feature expansion, which added the client-side file-encryption tool [62], the ML-DSA-87 authenticity layer that signs each served public key, role-based access control, the tamper-evident audit log, and multi-factor authentication. What the discipline buys is diagnosis. A failure during development had to live in the just-added layer, because everything underneath had already been proven; problems arrived one at a time in a known place instead of dispersed through untested code, which is what made the failures described below tractable at all. The governing rule: nothing counted as complete until tested against its intended behaviour, and for cryptographic code that meant an actual encrypt-decrypt round trip, never a check that it merely ran. That rule sounds obvious and is routinely broken; holding to it is most of what the discipline means in practice. The authenticity layer added in this phase followed the same pluggable, self-verifying backend pattern established for the key encapsulation, so that its signature backend is likewise proven with a real sign-and-verify round trip at initialisation before it is trusted. Its pure-Python fallback produces ML-DSA-87 signatures that are fully interoperable with the native library, each verifying against the other under the same public key.

### B. Design Revisions Forced by Implementation

Two decisions at the centre of the final architecture began as failures. Both are worth recording because each shows how building an artefact yields knowledge that paper design would not have produced.

The first involves the post-quantum backend. The plan treated the Open Quantum Safe liboqs library as an ordinary dependency, and it is not one: its Python binding ships no compiled C library but builds it on first import, using a C toolchain the development environment did not have, so the dependency could not be satisfied precisely where it most needed testing. Worse, the failure was silent. In certain states, the import succeeded, a naive presence check passed, and the backend only failed when actually asked to encapsulate. The resolution turned the fault into a feature. A pure-Python fallback backend was added, resting on a FIPS 203 implementation that needs no native build and producing the identical ML-KEM-1024 wire format, so the two backends are genuinely interchangeable. Selection was then made self-verifying: before any backend is declared usable, it must perform a real key generation and a full round trip at initialisation, and if it cannot, the system falls back to the portable implementation. The general lesson is uncomfortable and useful. A dependency that should have worked did not; its failure mode defeated the standard check, and the durable fix was a backend that proves itself before it is trusted.

The second involves where Sentinel Gate, the abuse-prevention gateway, actually lives. The original design placed it as a separate network service in a reverse-proxy chain: TLS terminates at the proxy, the proxy forwards to the gateway, the gateway forwards to the key service. As designed, it could not be built. The gateway had no mechanism for forwarding requests onward because it was architected to wrap an application mounted inside it, not to proxy to an external process. Rather than bolt on a forwarding layer, the topology was reversed: the gateway became in-process middleware inside the key service itself. A request now passes through the proxy, then kernel-level filtering, then the gateway middleware, which performs rate limiting, reputation scoring, and anomaly detection, and only then reaches the key-service handler, all inside one process. The reversal turned out stronger on its own merits. One network hop gone; one service fewer to run and monitor; one shared secret that no longer has to pass between two processes; and the gateway sees the full request context directly instead of a forwarded summary. The two components stay separate codebases with separate test suites and meet only at deployment: two artifacts on disk, one process at runtime. One consequence needs stating. The embedded gateway reads the client address from the proxy's forwarded header, and the deployment trusts that header only when the request arrives from the proxy on the local host, so clients are identified correctly. An attacker gains nothing by sending a forwarded header of their own.

### C. Ensuring Correctness Under Concurrency

One correctness problem shaped the gateway's rate limiter, and it is recorded here because the fix was a decision, not a default. A limiter that reads a counter, decides, and writes back races it. Under concurrent load, two requests can both read a permissive state before either writes, so both get in, and the limit fails exactly when parallel load is heaviest, which is exactly when it matters. The window was closed at the storage layer. The limiter's state transition runs as a single indivisible script inside the in-memory data store: read the token bucket, test it, decrement it, return the decision, with no concurrent request able to interleave, while the higher-level scoring logic stays in the application. The same atomic mechanism maintains the per-identity feature vector the anomaly detector uses. Testing this needed a way to exercise atomic, concurrent behaviour without a live data-store instance; an in-process emulation supporting the same scripting interface supplied it, so the limiter was validated alone before anything was built on top, in keeping with the bottom-up rule above.

### D. Empirical Evaluation Strategy

The evaluation, reported in full later, was built to answer the feasibility questions the contributions raise, and it deliberately keeps two kinds of measurement apart. The first isolates the post-quantum primitive on constrained embedded hardware to establish a floor: what is the least machine that can run this? An ARM Cortex-M4 microcontroller ran the standard pqm4 implementation of ML-KEM-1024, clocked low with zero flash wait states. Hence, cycle counts come out deterministic and comparable to published figures, timed by the on-chip cycle counter, with peripheral output excluded from the timed region so the number measures the cryptography and not the cost of reporting it. The second measures the complete system against live deployment across a wide range of file sizes and several hardware configurations, including a wide-area network path, because that is the behaviour an adopter actually experiences. Every software timing discards a warm-up run, reports the median or the mean with standard deviation over many iterations, and uses the production backend rather than the portable fallback, so the figures describe what an enterprise would deploy, not a laboratory convenience. The gateway was measured separately again: the key service placed under sustained abusive load from a distinct network origin, recording both the share of malicious traffic filtered and the latency a legitimate user pays during the attack. Keeping the three apart (embedded floor, system behaviour, abuse resistance) is deliberate. Each answers its own question, and blending them would blur all three; the microcontroller result carries the affordability claim, the system measurements the deployability claim, and the flood test the availability claim.

## V. Background and Preliminaries

Section II said what breaks; this section says what QCG uses instead, and why each choice resists the same attacks. The treatment stays brief on purpose. Every primitive here is standardised, and the security arguments live in the cited literature rather than in this paper.

### A. Classical Foundations and the Point of Failure

Internet security has rested on three primitive families. RSA draws its security from integer factorisation [25]. Elliptic-curve cryptography draws its own security from the elliptic-curve discrete logarithm problem [26] and reaches equivalent strength at far smaller keys than RSA. The Advanced Encryption Standard, a symmetric substitution-permutation network [28], [29], [51], does the bulk encryption. The first two are public-key schemes, the machinery of key establishment and signatures; the third is symmetric. And the fault line from Section II runs exactly between them: Shor solves factorisation and discrete logarithm in polynomial time and takes the asymmetric pair with it. At the same time, Grover manages only a square-root speed-up against key search, leaving AES-256 a wide residual margin. QCG replaces what falls on the broken side of that line and keeps what stands.

### B. Lattice-Based Cryptography and Module-LWE

Among the families NIST surveyed, lattices proved the most productive, supplying three of the four algorithms first selected [30]. Their hardness rests on problems like the shortest and closest vector problems, for which no efficient algorithm is known, classical or quantum; unlike factoring and discrete logarithm, they expose no abelian hidden-subgroup structure for Shor to exploit [31]. The modern line builds on Regev's Learning With Errors problem [32], the task of recovering a secret from noisy linear equations, which carries the rare property of a worst-case to average-case reduction: breaking random instances solves hard lattice problems in the worst case. QCG's primitives use the module variant, Module-LWE, which swaps integer entries for polynomial-ring elements and cuts key sizes by an order of magnitude without giving up the reduction [33]. ML-KEM, formerly CRYSTALS-Kyber, and ML-DSA, formerly CRYSTALS-Dilithium, both stand on Module-LWE. Their quantum resistance is not a proof, because cryptography cannot provide one. It is the absence, after decades of cryptanalysis, of any quantum algorithm that meaningfully beats classical lattice methods [34]. That absence is the strongest statement the field knows how to make.

### C. ML-KEM-1024 for Key Encapsulation

QCG protects its symmetric keys with ML-KEM-1024, the scheme formerly known as CRYSTALS-Kyber, standardised in FIPS 203 [35]. A key-encapsulation mechanism does not encrypt data. It lets two parties agree on a secret over a hostile channel: generate a keypair, encapsulate against the recipient's public key to produce a shared secret plus a ciphertext, and decapsulate with the secret key to recover the same secret. In QCG, that shared secret guards the AES-256 data key and nothing else. The standard offers three parameter sets, ML-KEM-512, ML-KEM-768, and ML-KEM-1024, at NIST categories 1, 3, and 5 [35]; QCG runs the last so that the KEM's category-5 strength matches AES-256 exactly. Size also argued for it. Public keys and ciphertexts come to roughly 1,568 bytes and secret keys to roughly 3,168 bytes [35], larger than their elliptic-curve counterparts, trivial for any ordinary client. And the scheme carries IND-CCA2 security, the strongest standard KEM notion, obtained through a Fujisaki-Okamoto transformation, so an adversary holding responses to many chosen ciphertexts still cannot recover the shared secret [35]. As Section II observed, that guarantee speaks to the algorithm; the implementation must earn its own.

### D. ML-DSA (Dilithium) for Authenticity

Confidentiality needs a partner. Digital signatures supply it, establishing that data came from the claimed sender and arrived unaltered; in QCG, a signature attests that encrypted material and its key data originated with the legitimate client rather than a substitute. The scheme is ML-DSA, formerly CRYSTALS-Dilithium, standardised in FIPS 204 [36]. It rests on Module-LWE together with the module short-integer-solution assumption, and it is deliberately easy to implement safely: uniform sampling only, with no floating-point arithmetic anywhere in signing, which is why NIST names it the default signature scheme. There is a policy angle too. Federal policy treats the authenticity half of the migration as the harder one, giving post-quantum authentication a deadline a full year behind key establishment [63], and deployed systems bear that judgement out, with post-quantum signing markedly behind post-quantum encryption in practice. QCG therefore builds the signature layer in as a first-class requirement, not an optional extra.

### E. AES-256-GCM and the KEM-DEM Construction

The third primitive encrypts the data itself. Grover's algorithm cuts an $n$-bit key's effective strength from $2^n$ to $2^{n/2}$ [19], which leaves AES-256 at an effective 128 bits against quantum search, beyond feasible reach and deliberately matched to ML-KEM-1024's category-5 strength so that neither primitive is the weaker link. Two decades of cryptanalysis without an efficient attack [28], [51] and hardware acceleration through AES-NI settle the rest of the case. QCG runs in Galois/Counter Mode, whose built-in authentication tag provides confidentiality and integrity in one pass [37]: tamper with the ciphertext, and decryption itself reports it, with no separate message-authentication code to manage.

The primitives meet in the KEM-DEM construction, the established way of building public-key encryption from a KEM and a data-encapsulation mechanism [38]. Each object's data is encrypted under a fresh AES-256 key in GCM mode; that key is encapsulated under the recipient's ML-KEM-1024 public key; decryption reverses the two steps. The ML-KEM shared secret never touches user data directly. It protects only the symmetric key, and this division of labour keeps the design efficient: data of any size is handled only by the fast symmetric cipher. At the same time, the asymmetric operation remains a fixed cost per object.

Why hybrid at all, rather than purely post-quantum? Risk management, not asymptotics. Lattice schemes are well studied, but their cryptanalytic history is a fraction of the length of AES's, and a little residual possibility of progress against Module-LWE has to be priced in [22]. The construction, to be precise, is a standard KEM-DEM composition and not two independent encryptions of the plaintext: ML-KEM-1024 protects the AES-256 key, AES-256-GCM protects the data. Against the harvest-now-decrypt-later adversary, its value rests on both primitives resisting a quantum attacker: AES-256 keeping its effective 128-bit margin under Grover, and ML-KEM-1024 being built for quantum resistance, so recorded ciphertext yields nothing unless the specific primitive guarding a stage is broken. The hybrid element is the pairing itself: post-quantum key establishment over a classical authenticated cipher, so an advance against the lattice assumption alone does not expose the plaintext, and a break of the symmetric cipher alone does not recover the key protecting it. Nothing in the design rests on a single new assumption being sound. That client-side defence-in-depth, sized to SME realities, is what separates QCG from the purely post-quantum and the purely classical alternatives alike.

### F. The KEM-DEM Hybrid as the Foundation of QCG

Layer 1 realises the paradigm directly [38]. For every encryption, ML-KEM-1024 encapsulation against the recipient's public key yields a fresh 256-bit shared secret and the encapsulation ciphertext in a single operation [35]; the plaintext goes under that secret with AES-256-GCM [37]; both ciphertexts travel or rest together; decryption reverses the process, as Fig. 2 shows. Three choices in that flow are deliberate. The two primitives are paired at a common category-5 strength, so neither is the weak link. GCM authenticates, so no separate MAC is needed. And every operation draws a fresh key, so each object's confidentiality stands alone. Just as deliberate is what the construction leaves out: secure delivery of the recipient public key, sender authentication, and multi-device access all belong to the second layer, the self-hosted QS-KMS of Section VII, which runs on an affordable VPS under SME control and provides keypair generation, encrypted key storage, token-based session access, and the integrated abuse-prevention gateway [39]. So QCG enters Section VI claiming no cryptographic invention. It claims a disciplined arrangement of standardised primitives, aimed at the gap the earlier sections mapped: software-only, client-side, hybrid, and holding its keys where no cloud provider can reach them.

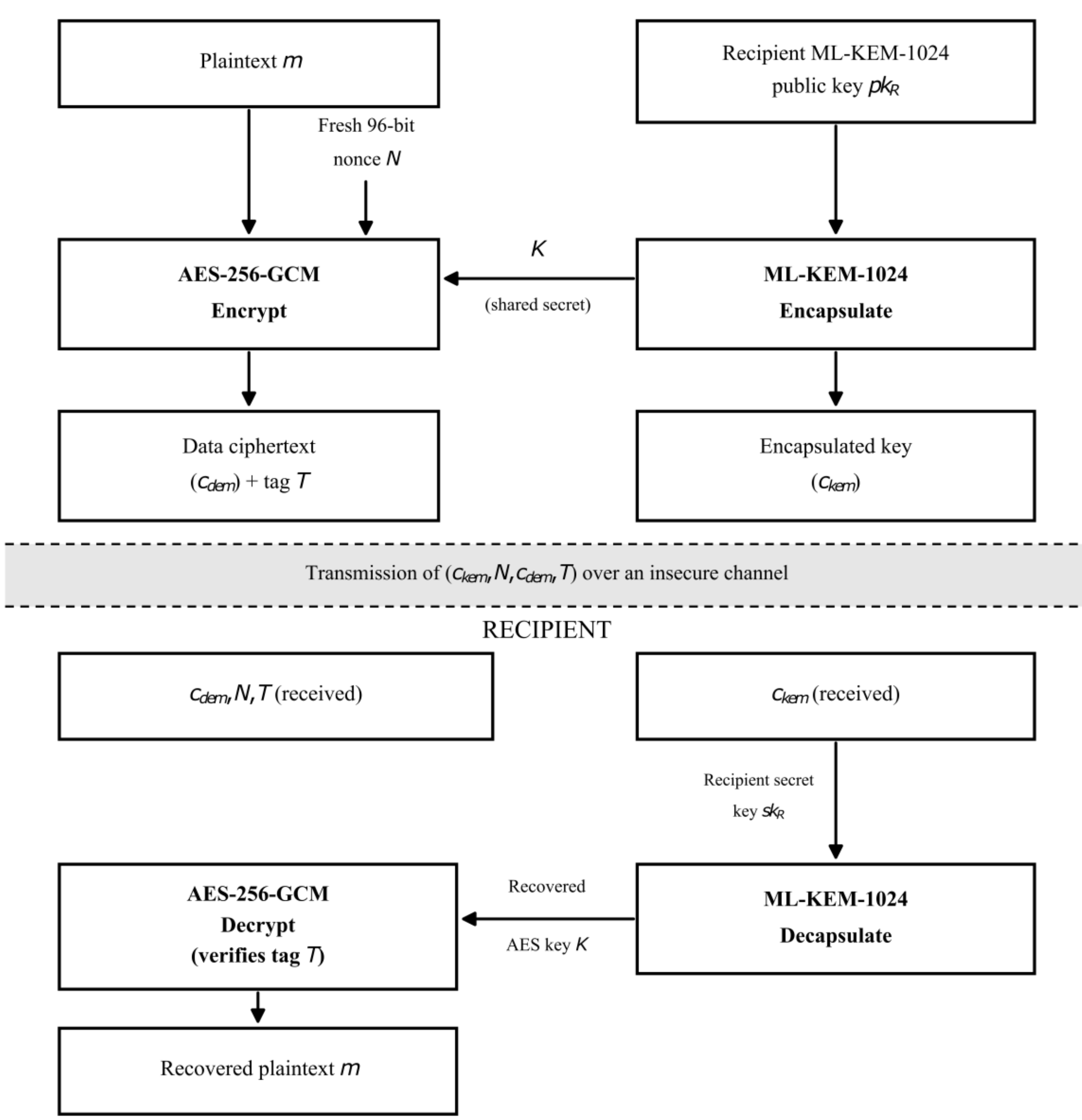

***Fig. 2.*** *QCG Layer 1 KEM-DEM data flow: plaintext m is encrypted under a fresh AES-256 key K to produce c_dem, and K is encapsulated under the recipient's ML-KEM-1024 public key pk_R to produce c_kem; the recipient recovers K using sk_R and decrypts c_dem. The public key pk_R enters this flow already authenticated, the client having verified its ML-DSA-87 signature under equation (11) before any encapsulation, so the signature operations themselves belong to Layer 2 and are deliberately outside the scope of this figure.*

## VI. QCG Framework Layer 1 - Client-Side Hybrid Encryption

### A. Overview of Layer 1

Layer 1 is where the data gets protected. It runs on the SME's own machine and does one thing: take a plaintext object and produce a ciphertext that resists classical and quantum attack alike, with the symmetric key that opens it sealed by a post-quantum mechanism. Everything in it executes before any byte leaves the machine. What it refuses to do is just as deliberate. Key management and authentication belong to Layer 2, described in Section VII, and keeping them out of Layer 1 keeps encryption independent of custody and exchange. The construction is the KEM-DEM framework of Section V made concrete: ML-KEM-1024, formerly CRYSTALS-Kyber, as the key-encapsulation mechanism, AES-256 in Galois/Counter Mode as the data-encapsulation mechanism. What follows covers the key material, the two protocols, the key lifecycle, and the defence against downgrade.

**B. Key Generation and Parameter Selection**

Layer 1 works with two kinds of key material: a long-term ML-KEM-1024 keypair belonging to the recipient, and a fresh symmetric secret drawn at every single encryption. The keypair is created once, when the QCG service initialises, following FIPS 203 [35]:

(1) $(pk_R, sk_R) \leftarrow$ ML-KEM-1024.KeyGen()

The public key pk_R goes to any sender who wants it; the secret key sk_R stays with the recipient and never reaches the cloud, and how it is stored and used is Layer 2's business. The parameter set is Category 5, NIST's highest, matching the effort required to break AES-256 [35], so the KEM and the cipher are equals by design. The symmetric secret is handled differently, and the difference matters. It is not generated and then wrapped. Because ML-KEM is a key-encapsulation mechanism, the secret and its ciphertext are born together in one encapsulation, and for ML-KEM-1024 that secret is exactly 256 bits, precisely what AES-256 wants, so no key-derivation step intervenes. A fresh secret per encryption means a different key per object, and that gives per-object key independence: recovering the key to one object reveals nothing about any other. The claim is about isolation between stored objects; it is distinct from forward secrecy in the interactive-session sense, which is not claimed here. All randomness, keypair and per-object alike, comes from a cryptographically secure source per FIPS 203 [35] and NIST SP 800-90A [40], because a weak source would quietly hollow out the layer no matter how strong ML-KEM and AES remain on paper. Table III summarises the parameters.

**TABLE III. QCG CRYPTOGRAPHIC COMPONENTS, PARAMETERS, AND STANDARDS**

| Component | Algorithm or Value | Size | Standard |
|---|---|---|---|
| Key encapsulation | ML-KEM-1024 (Category 5) | - | FIPS 203 |
| Public key pk_R | - | 1,568 bytes | FIPS 203 |
| Secret key sk_R | - | 3,168 bytes | FIPS 203 |
| Encapsulation ciphertext c_kem | - | 1,568 bytes | FIPS 203 |
| Shared secret K | - | 256 bits | FIPS 203 |
| Data encryption | AES-256-GCM | 256-bit key | FIPS 197, SP 800-38D |
| Nonce N | - | 96 bits | SP 800-38D |
| Authentication tag T | - | 128 bits | SP 800-38D |
| Digital signature | ML-DSA-87 (Category 5) | 4,627 bytes (sig) | FIPS 204 |

**C. The Encryption Protocol**

Encrypting a plaintext object m takes four steps, and the sender needs exactly one thing: the recipient's public key pk_R.

**Step 1. Establishing the key.** The sender runs ML-KEM encapsulation against pk_R and receives two things at once: a fresh 256-bit shared secret K and the ciphertext c_kem that carries it.

(2) $(c_{kem}, K) \leftarrow$ ML-KEM-1024.Encaps$(pk_R)$

**Step 2. Preparing the nonce.** A fresh random 96-bit nonce N is drawn for AES-256-GCM. A nonce must never repeat under one key, and since K is used exactly once, that is already guaranteed; the random draw follows NIST SP 800-38D [37] regardless.

**Step 3. Encrypting the data.** The plaintext m goes under K with AES-256-GCM, which returns the data ciphertext c_dem and a 128-bit authentication tag T.

(3) $(c_{dem}, T) \leftarrow \text{AES-256-GCM.Enc}(K, N, m)$

**Step 4. Assembling the output and clearing the key.** The protected object is the tuple (c_kem, N, c_dem, T), and it goes to the cloud. K is then wiped from client memory; the sender has no further use for it, and keeping it around would only widen the attack surface.

K never touches a disk and never crosses a wire. The only element that leaves the client is c_kem, and only the holder of sk_R can open it, so what sits in the cloud is safe on both counts: neither the data nor the key is available to anyone but the holder of the recipient's secret key.

### D. The Decryption Protocol

Decryption reverses the path and needs the recipient's secret key sk_R, supplied through the Layer 2 service of Section VII. Starting from a protected object (c_kem, N, c_dem, T), the recipient proceeds as follows.

**Step 1. Recovering the key.** c_kem and sk_R go into ML-KEM decapsulation, which returns the same shared secret K that encryption produced.

(4) $K \leftarrow \text{ML-KEM-1024.Decaps}(sk_R, c_{kem})$

**Step 2. Decrypting and verifying together.** The recipient decrypts c_dem under K and N, and the very same operation checks the tag T.

(5) $m \leftarrow \text{AES-256-GCM.Dec}(K, N, c_{dem}, T)$

**Step 3. Handling the outcome.** GCM verifies the tag before it releases a single byte of plaintext. A mismatch, whether from a transmission error or deliberate tampering with c_dem, N, or T, fails the whole operation: QCG discards the partial result and records the event. Only a valid tag releases m.

**Step 4. Clearing the key.** As on the sending side, K is erased from memory the moment the operation ends, success or failure.

The authenticated mode is the point, not a convenience. GCM binds the tag to the exact ciphertext, so almost any change to a stored object is caught, and the whole family of attacks that edit ciphertext hoping for a predictable change in the plaintext dies there. What survives tampering is only a log entry telling the SME it happened.

### E. Key Lifecycle Management

How long Layer 1 stays secure depends on more than the algorithms; it depends on how keys are treated across their whole life, and QCG follows the six-stage lifecycle of NIST SP 800-57 [27]. Generation: the one-time keypair is made at setup, and a fresh symmetric secret at every encryption, both from a secure entropy source. Storage: sk_R exists only in encrypted form inside the Layer 2 service, never in plaintext and never in the cloud, while session secrets live in client memory and nowhere else. Distribution: pk_R travels freely because it can; secret keys never travel at all, and decapsulation happens inside the key service so sk_R does not move between devices. Rotation: the symmetric side rotates itself, since every encryption draws a new secret, and the long-term keypair rotates at whatever frequency the SME chooses, provided a retired secret key is kept to open the objects sealed under it. Revocation closes a gap common to client-side schemes: a key judged compromised is marked in the key service, and every further decapsulation with it is refused. Confidentiality rests entirely on sk_R, so revocation is the SME's first response to compromise and has to be designed in from the start, not bolted on after. Destruction ends the story: session secrets are cleared the instant they are used, retired long-term keys are destroyed once they protect nothing, and from that moment the associated data is unrecoverable, a finality the SME must weigh against its data-retention obligations.

### F. Downgrade Attack Mitigation

Section II warned that any system mixing classical and post-quantum algorithms invites downgrade attacks, where an adversary leans on the negotiation until both sides settle for the weaker mode. QCG's answer is not to detect the interference. It is to make the interference pointless. There are no classical key-establishment alternatives inside QCG at all: no list of algorithms to pick from, no RSA or elliptic-curve fallback, no legacy mode kept alive for compatibility. The downgrade path that exists in hybrid negotiation-based systems is not present here. A downgrade attack works by forcing negotiation towards a weaker option, and where the suite offers no weaker option, there is nothing to force.

Four rules, enforced on every client, hold the property in place. The only key-establishment method in the system is ML-KEM-1024 encapsulation; no classical alternative exists. The only data-encryption mechanism is AES-256-GCM; no shorter key, weaker cipher, or unauthenticated mode is accepted. Any object that fails to match the pattern (c_kem, N, c_dem, T) under exactly these algorithms is rejected, not processed. And the signature mechanism for authenticity, set out in Section V, is fixed to a post-quantum scheme with no classical option beside it. Stripping out negotiation turns downgrade resistance into a property of the protocol rather than something to watch for at run time, because the weaker suite an adversary would push does not exist to be pushed. It also happens to be the simpler system for the SME to run: one known suite, fewer knobs, and none of the misconfigurations that quietly weaken security.

## VII. QCG Framework Layer 2 - Self-Hosted Quantum-Safe Key Infrastructure

### A. Overview of Layer 2

Layer 1 protects data but cannot stand alone. It needs an ML-KEM-1024 keypair, and that raises three questions that Section VI deliberately left open: where the secret key lives, how a legitimate user reaches it from whatever device they happen to be on, and how the infrastructure holding it stays up and defended.

The central decision of Layer 2 answers all three at once: the key infrastructure is hosted and controlled entirely by the SME, and the whole trust model hangs on that. The public cloud, whether AWS, Google Cloud, or Azure, serves as nothing more than the untrusted host of encrypted bytes, while the keys live on an inexpensive virtual private server the SME alone manages. The distinction is not the technical one between a VPS and a cloud host. It is control: who holds the credentials, who sets the policies, and who has administrative access; in QCG, the answer to all three is the SME. The framework trusts the key infrastructure; it never trusts the storage layer at all.

Three components make up the layer. The Quantum-Safe Key Management Service, the QS-KMS, holds the key material, authenticates users, and decapsulates on request. Sentinel Gate, the Layer 7 abuse-prevention gateway, runs as in-process middleware inside the QS-KMS, so every request is inspected before any key-service logic ever sees it [39]. Redis, an in-memory store, keeps the gateway's state and the session-token record. The user sees none of it. They deal with the QCG client application, which bundles Section VI's encryption with the Layer 2 communication, and to them the whole thing looks like an ordinary file-encryption utility. Fig. 3 shows the three components and the traffic between them.

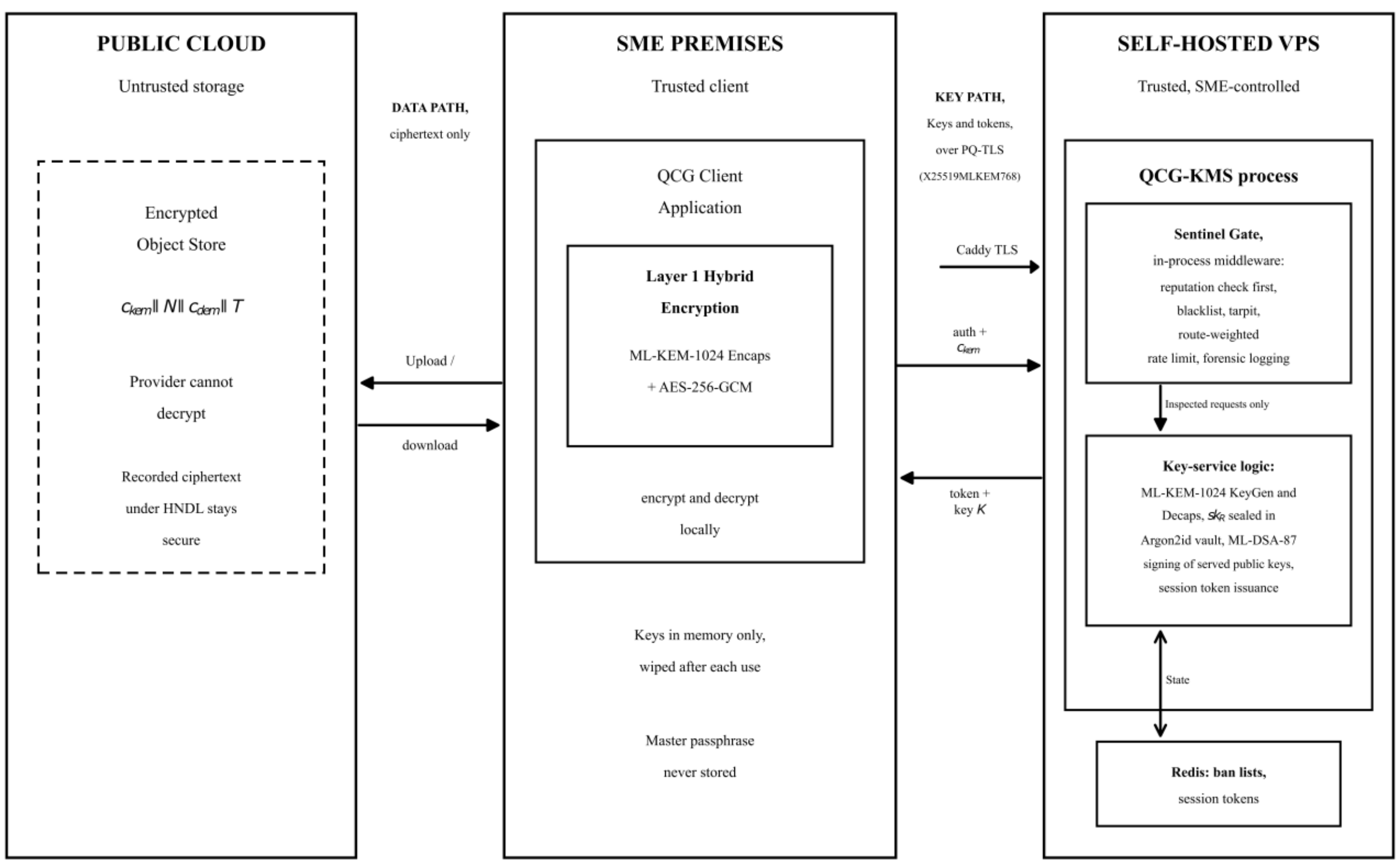

***Fig. 3.*** *QCG three-layer architecture: ciphertext is stored in the untrusted cloud (left), encrypted and decrypted locally by the QCG client (centre), and keys are managed by a self-hosted QS-KMS whose requests pass through the embedded Sentinel Gate abuse-prevention layer (right).*

**B. Architecture of the QS-KMS**

The QS-KMS follows established key-management principles [52] and is deliberately small, because a narrow service is a narrow target. It performs five functions: (i) Generation of the ML-KEM-1024 keypair during startup, (ii) Safekeeping of the secret key, (iii) User authentication, (iv) Decapsulation on demand by an authenticated user, and (v) Signing of every public key it serves with ML-DSA-87, so that a client can verify authenticity before use.

Initialisation never writes the secret key sk_R to the file system. A vault key comes first, derived from the user's master passphrase with Argon2id [41], the memory-hard password-based key derivation function of RFC 9106:

(6) $vk \leftarrow \text{Argon2id}(passphrase, salt, t, m, p)$,

where salt is random and stored with the vault, and t, m, and p are the time, memory, and parallelism parameters, set high enough to make brute force a losing proposition. The secret key is then sealed under the vault key with AES-256-GCM before it is stored,

(7) $C_{vault} \leftarrow \text{AES-256-GCM.Enc}(vk, N_v, sk_R)$,

so what the VPS disk actually holds is C_vault, never sk_R. Argon2id is slow and memory-hungry on purpose: an attacker who steals C_vault faces an impractical search over passphrases, and since the master passphrase is stored nowhere at all, the server contains no secret from which the vault key could ever be recovered.

Authentication falls straight out of the scheme: the user presents the passphrase over TLS. That handshake is itself a hybrid post-quantum exchange, because a classical handshake would hand the authentication traffic to exactly the harvest-now-decrypt-later collector the rest of the design is built against. The live deployment negotiates the X25519MLKEM768 hybrid group, verified against the running server [61], so the channel is quantum-safe independently of every application-layer protection this paper describes. The QS-KMS checks the passphrase against a stored Argon2id verifier; on success, it derives the vault key, decrypts sk_R into volatile memory, and opens the session. The secret key lives in volatile memory, and only there, for exactly as long as the session does.

An open session turns the QS-KMS into what cryptography calls a decapsulation oracle: it will decapsulate any c_kem an authenticated caller submits. That sounds like the weak point; it is the right question for a reviewer to press, and it deserves a straight answer. ML-KEM-1024 is IND-CCA2 secure, the adaptive chosen-ciphertext property Section V introduced, which means an adversary can feed the server chosen ciphertexts and study its behaviour without learning the secret key or anything useful about other ciphertexts. The property ML-KEM was designed to provide is precisely the property that makes server-side decapsulation safe to offer. What remains is operational rather than cryptographic: the risk that a stolen session token lets an attacker use the oracle as if entitled to it. That risk is governed by the token protections and access controls of Subsections C and D, and it assumes throughout that the user's endpoint is uncompromised, the boundary Subsection G examines.

While a session is open, sk_R sits in the VPS memory, and an adversary who controls the VPS could read it. That is tolerable for one reason only: the VPS belongs to the SME's own trusted infrastructure, not to the public cloud. An SME with stricter requirements can run the QS-KMS on physical hardware of its own instead of a rented VPS.

The API is as small as the service: authentication, decapsulation, revocation, a heartbeat for keep-alive, a health endpoint, and nothing else. Every one of them sits behind Sentinel Gate, so no request reaches QS-KMS logic without clearing the gateway first. And because C_vault is encrypted under a key derived from the passphrase, the vault can be backed up freely without exposing sk_R, which gives the SME a route back from losing the VPS altogether.

### C. Sentinel Gate as the Access Protection Layer

A self-hosted key service enjoys none of the abuse absorption a hyperscaler provides, and its endpoints are expensive to serve by design: Argon2id burns memory and time on purpose, and ML-KEM decapsulation is real computation. When every request triggers a heavy operation, whatever its outcome, an attacker needs very little traffic to impose real cost. That is why every request goes through Sentinel Gate [39].

As a Layer 7 gateway, Sentinel Gate reads application-layer information rather than counting packets, and each request clears its checks before any key-service logic runs. The checks are ordered by design. First comes an instantaneous reputation lookup against a Redis blacklist, which rejects a known-bad source within microseconds, before a single cycle of Argon2id is spent on it. A gateway that served first and checked afterwards would fail exactly here; Sentinel Gate runs its cheapest check first, before anything expensive starts.

A request that survives the lookup then passes an identity check, a quota check, and volumetric weighting before it reaches the QS-KMS, and every served request is logged for forensics. Volumetric weighting matches the defence to the cost an attacker is trying to impose. Each endpoint carries a weight reflecting what it costs to serve, so a campaign against the expensive endpoints exhausts its quota fastest. Table IV lists the weights.

**TABLE IV. KEY-SERVICE ROUTES AND ABUSE-CONTROL WEIGHTS**

| Endpoint | Function | Relative cost to serve | Assigned weight |
|---|---|---|---|
| /auth | Passphrase verification and vault decryption | High (Argon2id plus AES) | 10 |
| /decapsulate | ML-KEM decapsulation under an open session | Moderate | 5 |
| /revoke | Session token invalidation | Low (single Redis write) | 1 |
| /heartbeat | Session keepalive | Minimal | 1 |
| /health | Service status check | Minimal | 1 |

| Endpoint | Function | Relative cost to serve | Assigned weight |
| --- | --- | --- | --- |
| unrecognised | Any route not on the list above | Maximum penalty | 15 |

Four tactics do the defensive work. Dynamic blacklisting bars a source once it crosses an abuse threshold and records the decision in Redis. Progressive tarpitting adds delays that grow as a source approaches its limit, which frustrates automation without punishing a legitimate heavy user. Route-weighted rate limiting caps requests per source as just described. Forensic logging writes every event down for later analysis.

Because the gateway guards the most critical piece of the system, five hardening decisions go beyond its defaults. A trusted-device list exempts the SME's own machines from banning and tarpitting, while still logging them, so the organisation can never be locked out by its own protection. The fail mode is closed: if Redis is unreachable, the service denies rather than serves, because what it protects is the most sensitive data in the framework. Token revocation is synchronous, never queued, so revocation means now. Forensic logs are forwarded to a separate secured store, so compromising the KMS host does not also compromise the record of the compromise. And the bundled attack simulator stays public, which we regard as transparency rather than risk.

Redis is the right store for all of this. The checks must finish in milliseconds, or they become the latency they exist to prevent, and only an in-memory store with built-in expiry manages that: a disk-backed database is far too slow for a check that runs on every request, and a plain in-process structure would not survive across processes [42].

### D. Session Management and Token Lifecycle

An authenticated user needs to make further requests without re-entering the passphrase every time, and the system needs a way to end that access cleanly the instant it should end. Session tokens answer both. On authentication, the user receives a token drawn from a cryptographically secure pseudorandom number generator.

(8) $token \leftarrow \text{CSPRNG}(256)$

The token lives only in the client's volatile memory, never on disk, so when the session ends the machine keeps nothing of it. Every later request presents it to Sentinel Gate, which checks it against the Redis session store before anything goes through.

Three independent mechanisms bound the token's life, and the work is done by their combination, not by any one alone. The first is a hard expiry: a token lives one hour and then dies, active or not, which caps what a leaked token could ever be worth. The second is revocation on close: the QCG client sends a revocation request as it shuts down, and because revocation is synchronous in Redis, as the previous subsection explained, the token is dead the moment the client closes rather than up to an hour later. The third is a heartbeat: an active client checks in at a regular interval, and if the QS-KMS stops hearing it, because the client crashed or was closed carelessly, the token expires after a few missed beats. Together, the three cover the ordinary case, a user closing the software, and the untidy ones, a crash or a token quietly ageing out. A user can hold several sessions at once, one at the office and another at home, each under its own independent token.

### E. End-to-End Workflow

The subsections above describe the parts in isolation. This one assembles them, following three realistic scenarios, so a cryptographically literate reader can watch the guarantees hold step by step and a business owner can recognise their own working day in it.

Scenario 1, first-time setup. The owner of a small architectural firm provisions an inexpensive virtual private server, personally or through an IT contractor, installs the QS-KMS, Sentinel Gate, and Redis on it, and puts the QCG client on the main office computer. Setup asks for a master passphrase, which the owner keeps nowhere but in memory. Underneath, QCG generates an ML-KEM-1024 keypair (Section VI), derives a vault key from the passphrase with Argon2id, seals the secret key under it, and sends the encrypted vault and the public key to the QS-KMS over a secure channel, registering the workstation as a trusted device so the firm's own machines are never blocked by their own defences. The firm now has a working installation, a known public key for encryption, and a secret key sealed in the vault. From here, the owner has one thing to remember: the passphrase.

Scenario 2: an ordinary working day. Weeks later, an architect saves drawings into the QCG-protected folder, which synchronises to cloud storage. To the architect, nothing happened. Underneath, a full Layer 1 cycle ran: the client encapsulated against the firm's public key to get a fresh key and its encapsulation, encrypted the drawings under that key with AES-256-GCM, bundled encapsulation and ciphertext, wiped the key from memory, and uploaded the result. The drawings now sit in the cloud unreadable to any current adversary and to a future quantum one. The QS-KMS was never contacted, because encryption needs only the public key already on the client; the whole operation was local.

Scenario 3, opening a file elsewhere. That evening the architect works from home on a laptop that has never seen QCG, so the client is installed and the master passphrase entered, and it connects to the firm's QS-KMS over a secure channel. The request meets Sentinel Gate first, which checks the source against the blacklist and the rate limits before anything is forwarded. The QS-KMS validates the passphrase against the Argon2id verifier, derives the vault key, briefly decrypts the secret key into memory, and issues a one-hour session token. The client downloads the encrypted drawings and sends the encapsulation with the token; the QS-KMS decapsulates server-side and returns the key for local decryption. When the architect closes the file, the client sends a revocation, the token dies in Redis, the secret key leaves the QS-KMS memory, and the laptop retains nothing capable of decrypting anything. Access lasted exactly as long as it was needed, and the secret key never once touched the architect's device.

### F. Deployment Considerations

A framework is only useful if its target organisations can actually run it. Layer 2 asks for little: the QS-KMS, Sentinel Gate, and Redis share a small VPS with two virtual processors and two gigabytes of memory, on offer from several providers at four to six euros a month, and free on some tiers. A complete quantum-safe key infrastructure therefore costs an SME a few euros a month, or nothing.

Some configurations are not optional, and each follows from an earlier section: Redis must persist its ban and session lists so state survives a restart; Sentinel Gate must fail closed, so a Redis outage denies access instead of disabling protection; and client-to-QS-KMS traffic runs over hybrid post-quantum TLS, the live deployment negotiating X25519MLKEM768 at the transport layer.

The limitations deserve naming too. In the basic installation, the QS-KMS is a single point of failure: while it is down, encrypted files cannot be read, though they stay secure and encryption keeps working, and an SME that needs more can run a second instance at extra cost and complexity. The passphrase model has its own sharp edge. The secret key can be recovered only through the master passphrase; the passphrase is stored nowhere, and so forgetting it makes the data permanently unrecoverable; that is the price of a system nobody but its user can open. And the whole design assumes the endpoint itself is trustworthy, which is where the next subsection draws the boundary.

### G. Attack Mitigation Coverage

Section II raised the threats; a framework that raises threats owes each one a concrete answer. Table V maps every threat to the mechanism that answers it and the layer where that mechanism lives.

**TABLE V. THREAT-TO-MECHANISM MAPPING ACROSS QCG LAYERS**

| Threat (from Section II) | Mechanism in QCG | Where it lives |
|---|---|---|
| Harvest Now, Decrypt Later | ML-KEM-1024 plus AES-256-GCM hybrid encryption | Layer 1 |
| Man-in-the-Middle | ML-DSA-87 signed public keys plus post-quantum TLS | Layers 1 and 2 |
| Downgrade attack | No classical fallback path, algorithm lock | Layer 1 |

| Threat (from Section II) | Mechanism in QCG | Where it lives |
|---|---|---|
| DoS against key infrastructure | Sentinel Gate, reputation-first handling | Layer 2 |
| Brute-force authentication | Argon2id plus Sentinel Gate rate limiting | Layer 2 |
| Side-channel | Constant-time discipline in the liboqs implementation for timing channels; power and electromagnetic channels are implementation-specific and not evaluated here | Implementation |
| Endpoint compromise | Out of scope, stated as a boundary | Acknowledged limitation |

One entry in the table is deliberately qualified. Timing-channel resistance comes from the constant-time discipline of the liboqs implementation. Power and electromagnetic side-channels depend on the physical deployment environment rather than the architecture, and evaluating them sits outside the scope of this work.

Each mechanism in the table is examined in the security analysis of Section XI, which states the threat model explicitly and argues the construction's confidentiality by reduction to the security of its standardised components. With the design complete, the paper turns from what QCG is to how it was built and how it behaves under measurement.

# VIII. Implementation

## A. Implementation Overview

This section turns the framework of Sections V and VI into running code and shows it working on off-the-shelf hardware with free, open-source components throughout. Its job is to prove feasibility and to feed the evaluation of Section IX; the extra hardening a production deployment would want is held for Section X.

Python 3 carries the reference implementation, chosen for readable code and mature cryptographic libraries. The post-quantum operations, ML-KEM-1024 and ML-DSA-87, come from liboqs through its Python bindings [44], [45]. AES-256-GCM comes from pyca/cryptography. The QS-KMS is a FastAPI web service [43]. Argon2id from argon2-cffi seals the secret-key vault at rest per RFC 9106 [41], Redis holds session and gateway state [42] through redis-py, and Sentinel Gate [39] runs as in-process middleware, inspecting every request before the key-service logic sees it. Nothing in the stack costs money, and nothing needs special hardware, so the whole system runs on infrastructure any SME already has access to.

The code lives in two public repositories. The QCG proof-of-concept repository holds the client encryption module, the QS-KMS service, deployment settings, and the test scripts [46]. Sentinel Gate is maintained separately [39] because, although it is a modular component of QCG, it works as a general Layer 7 gateway in its own right. Both are public for one reason: so the results in this paper can be replicated by anyone.

## B. Layer 1 Implementation

Layer 1 is a client-side Python module on the SME machine with two operations, encrypt and decrypt, both entirely local. Encryption makes no network call at all, so routine file protection runs at local speed with no dependence on the key infrastructure, exactly as Section VII argued.

Before any encryption takes place, the client establishes that the recipient public key it holds is authentic. When pk_R is first retrieved from the QS-KMS, it arrives together with an ML-DSA-87 signature produced by the service over that key. The client verifies this signature against the QS-KMS ML-DSA-87 public key, which was obtained and stored locally during setup, using the verification routine in liboqs [44], [45]. If the signature does not verify, the client treats the public key as untrusted, aborts, and performs no encryption. This check is what closes the public key substitution avenue described in Section V and is the client-side half of the man-in-the-middle mitigation summarised in Section VII-G.

With the public key verified, the client encapsulates. It calls the liboqs ML-KEM-1024 encapsulation routine against pk_R, which returns the encapsulation ciphertext c_kem together with the shared secret K, exactly as equation (2) of Section VI specifies. One property deserves restating because the implementation leans on it: K is produced by the encapsulation itself, not chosen beforehand. That is what a key-encapsulation mechanism is. ML-KEM-1024 outputs a uniformly random 256-bit shared secret which, per FIPS 203, is fit for direct use as an AES-256 key, and QCG uses it directly as the data-encryption key. Listing 1 shows the verification and encapsulation steps together.

```
# Verify recipient public key, then encapsulate
if not mldsa.verify(pk_R, pk_R_sig, kms_mldsa_pub):
    raise SecurityError("public key signature invalid; aborting")

kem = oqs.KeyEncapsulation("ML-KEM-1024")
c_kem, K = kem.encap_secret(pk_R)   # K is produced by the KEM
```

The plaintext then goes under K with AES-256 in Galois/Counter Mode, as in equation (3) of Section VI. A fresh 96-bit nonce is generated by os.urandom for every encryption, backed on all supported platforms by the operating system's cryptographically secure random source. This produces the ciphertext c_dem and the 128-bit tag T, and the client assembles the protected object (c_kem, N, c_dem, T) and writes it out for upload.

The moment the object is assembled, the client disposes of K, overwriting the reference and invoking garbage collection. An honest limitation belongs here. A managed-memory language like Python cannot guarantee that every copy of a sensitive value dies at once, because the runtime may hold internal copies of its own. The implementation keeps K's lifetime as short as it can and never writes it to disk. Still, a deployment with the strictest memory-hygiene requirements would write this routine in a language with direct memory control, C or Rust. That is a property of the prototype, not of the design, and Section X returns to it.

Decryption runs the path backwards. The client fetches the protected object from the cloud, sends c_kem with its session token to the QS-KMS, and receives the recovered shared secret K, as Subsection C describes. AES-256-GCM decryption then follows equation (5) of Section VI, and GCM checks the authentication tag before releasing a single byte. Hence, an object altered in storage or in transit fails outright and yields nothing. Only a verified tag releases the plaintext, and then K is disposed of again.

### C. Layer 2 QS-KMS Implementation

The QS-KMS is a FastAPI application [43] on Uvicorn, running on the self-hosted VPS. Its interface is least-privilege and role-gated: routes for first-run setup, session management, per-key access control, encapsulation and decapsulation, and administration, nothing more. Every route sits behind the integrated gateway, so no request touches application logic without clearing inspection, and access turns on the caller's role and per-key grants rather than one shared credential.

Protecting the secret key at rest is the most security-critical piece of the service, and the code does exactly what Section VII-B specifies. At setup, the ML-KEM-1024 keypair comes from the liboqs key generation routine [44], [45]. The secret key never reaches the disk in the clear. A vault key is derived first from the master passphrase with Argon2id via argon2-cffi, following equation (6) of Section VII, at a time cost of three iterations, a memory cost of sixty-four mebibytes, and a parallelism of four, the parameters RFC 9106 recommends for high-security use [41]. The secret key is then encrypted under that vault key with AES-256-GCM, following equation (7), and only the resulting ciphertext ever touches the VPS disk. Listing 2 shows the derivation with its parameters spelled out, so the security configuration is open to inspection.

```
from argon2.low_level import hash_secret_raw, Type

vault_key = hash_secret_raw(
    secret=passphrase.encode(), salt=salt,
    time_cost=3, memory_cost=65536, parallelism=4,
    hash_len=32, type=Type.ID            # Argon2id, RFC 9106
)
```

Public key authenticity is established on the service side at registration time. When pk_R is registered during setup, the QS-KMS signs it with its ML-DSA-87 secret key using liboqs [44], [45] and stores the public key together with its signature. The signature scheme used is ML-DSA-87, the ML-DSA parameter set at NIST security Category 5, chosen to match the Category 5 strength of ML-KEM-1024 so that key authentication is no weaker than the encapsulation it protects. When a client later requests the public key, the service returns the key and its signature together, and the client verifies it as described in Subsection B. This is the service-side half of the man-in-the-middle mitigation, stated formally:

(9) $(pk_S, sk_S) \leftarrow \text{ML-DSA-87.KeyGen}()$

(10) $\sigma \leftarrow \text{ML-DSA-87.Sign}(sk_S, pk_R)$

(11) $\text{accept/reject} \leftarrow \text{ML-DSA-87.Verify}(pk_S, pk_R, \sigma)$

Equation (9) is executed once at service initialisation, equation (10) at each registration of a public key, and equation (11) by the client before any encapsulation under a retrieved key.

The authentication endpoint takes the passphrase over TLS, checks it against a stored Argon2id verifier, and on success derives the vault key, decrypts the secret key into memory, and issues a session token. The decapsulation endpoint takes a c_kem and a token, validates the token against Redis through the gateway, decapsulates with the in-memory secret key, and returns the shared secret. The secret key exists in memory only while the session does, and is erased when it ends.

Session tokens follow Section VII-D to the letter. A token is a 256-bit value from a cryptographically secure source, drawn in Python with the secrets module as equation (8) of Section VII shows, and stored in Redis with a one-hour time-to-live that enforces the hard cap. When a client closes, its revocation is written to Redis synchronously, never queued, so the token is invalid the instant the client closes and no stale token lingers in a backlog. A heartbeat at a fixed interval keeps the session alive; when the heartbeats stop, which is what a crash looks like, the session expires after a few missed intervals. Listing 3 shows the token issuance and the synchronous Redis write.

```
import secrets

secrets.token_bytes(32)                          # 256-bit token
redis.set(f"qskms:session:{token.hex()}",
          user_id, ex=3600)                      # hard 1-hour TTL
# revocation on client closure is a synchronous delete
# redis.delete(f"qskms:session:{token.hex()}")
```

**D. Sentinel Gate Integration**

Sentinel Gate [39] runs as in-process middleware inside the QS-KMS process, so every incoming request passes its Layer 7 inspection before the service logic runs. A request admitted by the TLS-terminating reverse proxy, Caddy 2.11.4, traverses the gateway middleware, where rate limiting, reputation scoring, and anomaly checks run, and only then reaches the key-service handler, all inside one process on the server. The configuration applies the five hardening decisions of Section VII-C rather than the component defaults. The fail mode is closed, so losing Redis denies access instead of disabling protection. The trusted-device allowlist carries the fingerprints of the SME's registered machines, so the organisation's own devices are never tarpitted or banned by the system meant to protect them, while their activity is still logged. Revocations are written synchronously, and forensic logs are forwarded to a separate access-controlled store rather than staying on the gateway host.

The route weights configured in the gateway are exactly those of Table IV in Section VII: the expensive authentication endpoint at the highest legitimate weight, decapsulation at a moderate one, the lightweight revocation, heartbeat, and health endpoints at the lowest, and any unrecognised route at the maximum penalty. Using the same values in the implementation as in the paper is deliberate. The behaviour this paper describes is the behaviour a reader will find in the deployed system.

Sentinel Gate and the QS-KMS share one Redis instance on the VPS, which spares a small server the cost of running two stores. Their keys cannot collide because their namespaces are separated: gateway state, ban lists, and rate-limit counters live under a gateway prefix, while session tokens live under the service prefix shown in Listing 3. The two concerns stay isolated inside the shared store, and the contents of Redis stay easy to read during operation.

### E. Reproducibility and Open-Source Availability

Everything is public so that every result can be checked. The proof-of-concept, client module, and QS-KMS service together have their own repository [46].

Inside it, the key service and its web console live in QCG-KMS [46], the gateway in Sentinel-Gate-QCG [39], and the standalone command-line client built in QCG-CLI-Kit [62]. Every benchmark script and raw result file sits in Benchmarks, cited individually against the table it supports [54], [55], [56], [57], [58], [59], [60], [61], with a folder README mapping each file to its table. The repository README documents the whole deployment procedure, from provisioning the VPS through persistent Redis configuration, QS-KMS setup, Sentinel Gate configuration, and client installation. A reader with a cheap VPS and an ordinary computer can rebuild the entire system from the repository alone.

Cryptographic results depend on the exact software that produced them, so Table VI records the version of every major dependency [60]. With the versions pinned, the benchmarks of Section IX can be reproduced under the same conditions, and later changes in the underlying libraries cannot quietly redefine what the numbers mean.

**TABLE VI. DEPLOYED SOFTWARE STACK AND VERSIONS [60]**

| Component | Library | Version |
| --- | --- | --- |
| Post-quantum algorithms | liboqs | 0.15.0 |
| Python PQC bindings | liboqs-python | 0.15.0 |
| Symmetric encryption | pyca/cryptography | 46.0.3 |
| KMS web framework | FastAPI | 0.136.3 |
| ASGI server | Uvicorn | 0.49.0 |
| Password-based key derivation | argon2-cffi | 25.1.0 |
| Session and gateway store | Redis | 8.0.5 |
| Redis Python client | redis-py | 8.0.0 |
| TLS terminator | Caddy | 2.11.4 |
| Runtime | Python | 3.14.4 |
| Pure-Python ML-KEM fallback | kyber-py | 1.0.1 |
| Pure-Python ML-DSA fallback | dilithium-py | 1.4.0 |

## IX. Evaluation

The evaluation answers the three questions the contributions of Section I raise: whether the post-quantum key encapsulation is cheap enough for the modest hardware an SME owns, what overhead the full workflow imposes across a realistic range of file sizes on a live deployment, and how well the self-hosted key infrastructure holds up under sustained abusive load. Throughout, every software timing discards a warm-up run, reports the median or the mean with standard deviation over many iterations, and uses the native liboqs backend an SME would actually deploy.

### A. Experimental Methodology

Two classes of measurement are reported. The first benchmarks the post-quantum primitive alone on a microcontroller, characterising ML-KEM-1024 on hardware far weaker than any workstation. The second measures the system against the live deployment at qcgkms.cloud, a Hetzner CX22 virtual private server with two vCPUs and roughly 3.7 GB of memory, running the key service behind a TLS-terminating reverse proxy, Caddy 2.11.4, which negotiates the X25519MLKEM768 hybrid post-quantum key exchange by default; the remote client is an Intel Core i9-14900HX workstation in Karachi across a network path of roughly 175 to 185 ms round trip. The two stay separate because they answer different questions, and mixing them would blur both.

### B. Embedded Performance Floor: ML-KEM-1024 on ARM Cortex-M4

To fix a lower bound on the hardware that can run the primitive, ML-KEM-1024 was benchmarked on an STM32F407 [54], a constrained ARM Cortex-M4 device of the kind an SME might deploy at an embedded endpoint. The implementation comes from pqm4, the established Cortex-M4 benchmarking framework, so the figures compare directly with published ones. The device ran at 24 MHz for zero flash wait states and therefore deterministic cycle counts; timing came from the on-chip DWT cycle counter; and all serial output sat outside the timed region, so peripheral latency never touches the measurement.

Each of the three operations was measured over twenty runs, timed in milliseconds at full precision, with cycle counts derived by multiplying by twenty-four thousand, since at 24 MHz one millisecond is exactly twenty-four thousand cycles. Table VII gives the results; Fig. 4 shows the rig.

**TABLE VII. ML-KEM-1024 COST ON STM32F407 AT 24 MHZ [54]**

| Operation | Mean (ms) | Std. dev. (ms) | Derived mean (cycles) | Min (ms) | Max (ms) |
|---|---|---|---|---|---|
| KeyGen | 40.84 | 0.101 | 980,268 | 40.80 | 41.27 |
| Encapsulation | 41.46 | 0.101 | 994,920 | 41.41 | 41.88 |
| Decapsulation | 43.96 | 0.099 | 1,055,100 | 43.92 | 44.38 |

*pqm4 commit a24bb4b662016968c19f5e6a0719c9ad530f0286; arm-none-eabi-gcc 12.3.rel1; STM32F407G-DISC1 (MB997E), Cortex-M4 at 24 MHz, zero flash wait states; DWT cycle counter; serial output excluded from the timed region; twenty runs per operation.*

Standard deviations near a tenth of a millisecond indicate stable measurements. The reading is direct. If key encapsulation finishes in roughly forty milliseconds on a microcontroller of this class, then on any ordinary workstation, where the same operation takes well under a millisecond as the next subsection shows, the primitive is no barrier at all. This is the performance floor, and it carries the accessibility claim: the client-side cryptography needs no special hardware and runs comfortably on a device costing a few tens of dollars.

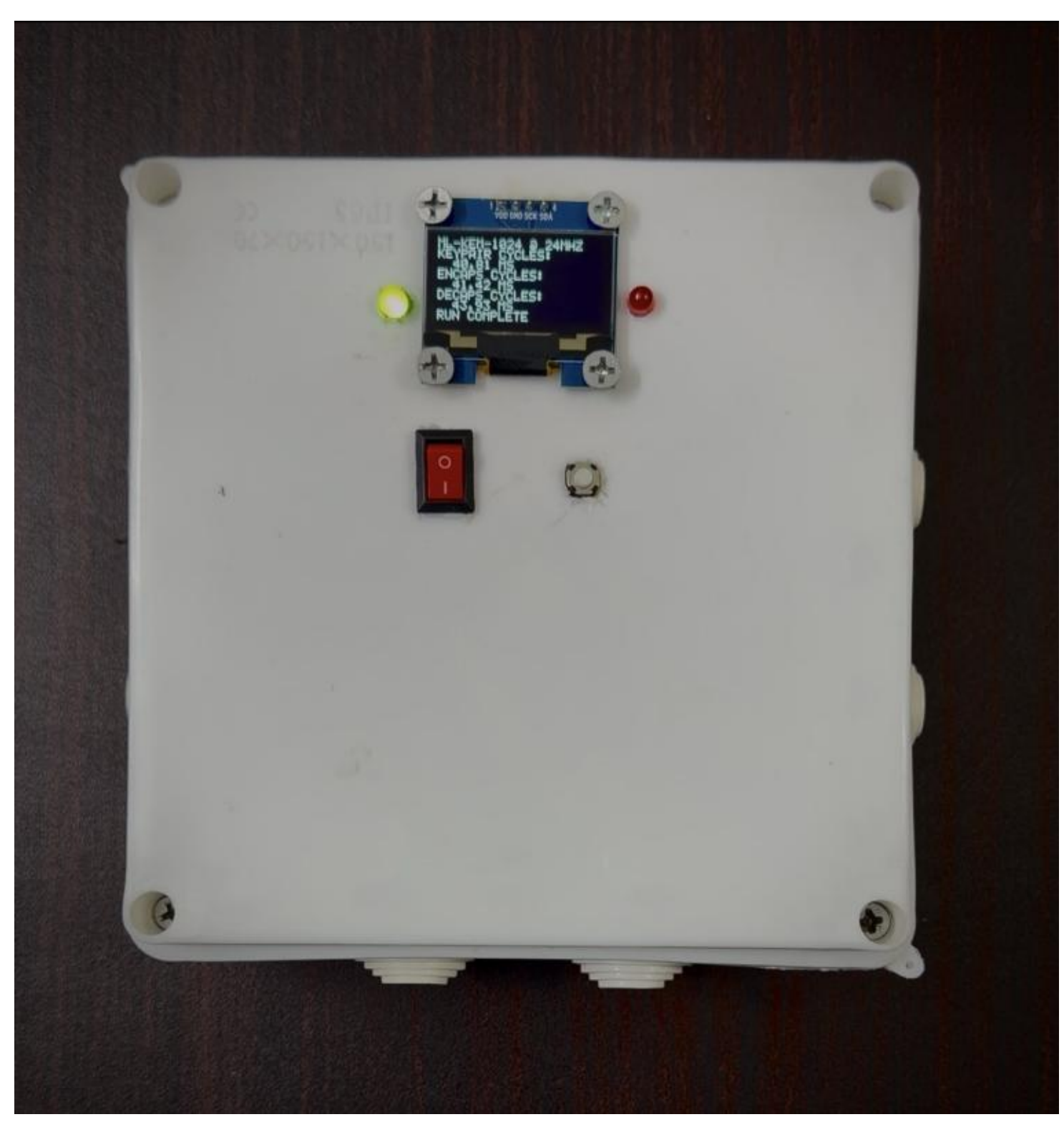

***Fig. 4.*** *STM32F407 embedded benchmark rig in its sealed enclosure, with the OLED reporting the measured ML-KEM-1024 key-generation, encapsulation, and decapsulation timings at 24 MHz and the status indicators showing run completion.*

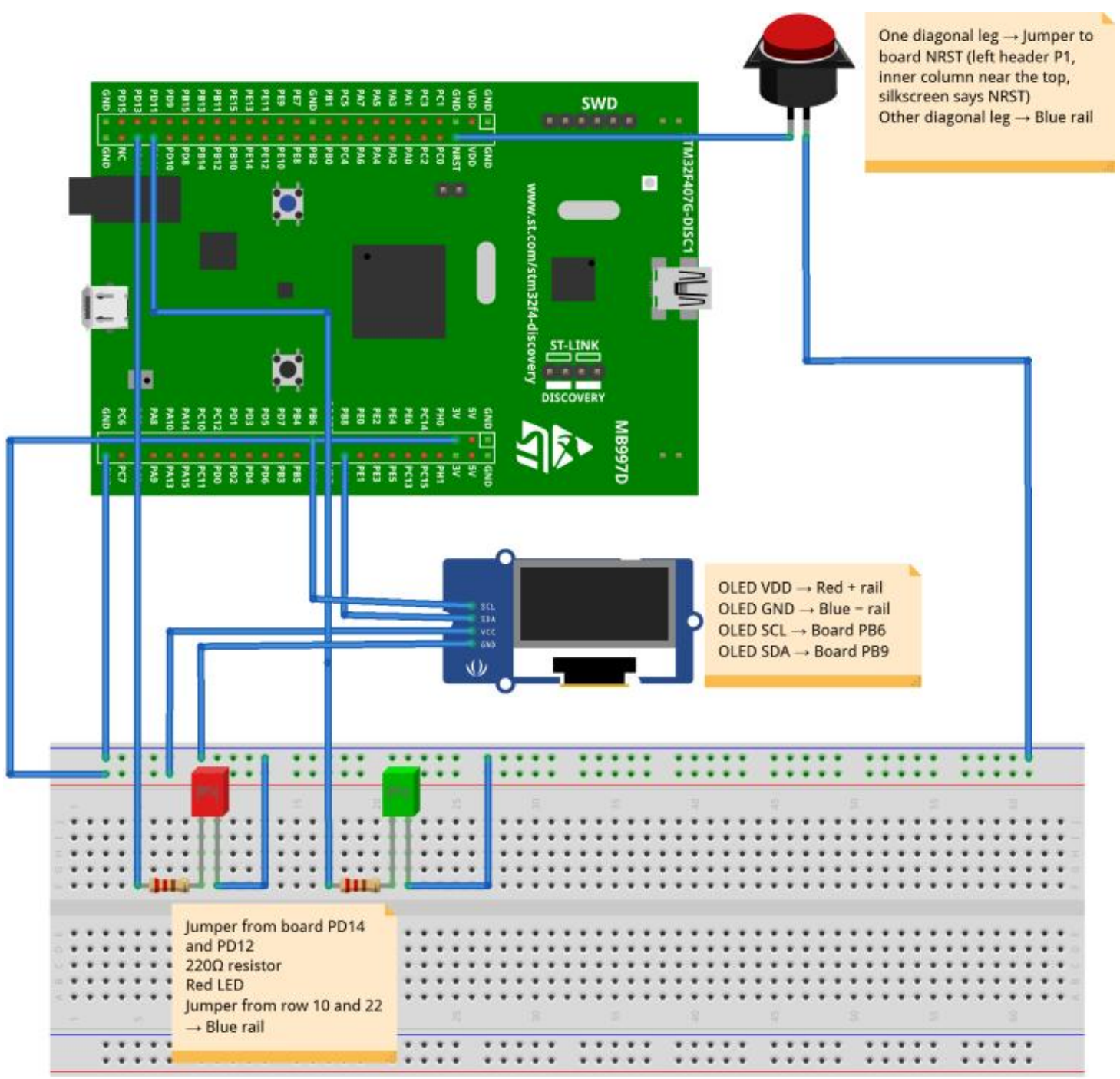

***Fig. 5.*** *Wiring of the STM32F407 benchmark rig: the I2C OLED on pins PB6 and PB9, the status LEDs on PD12 and PD14 through current-limiting resistors, and the reset control on NRST.*

### C. End-to-End System Performance

The full workflow was then measured on the live deployment in two configurations to separate computational cost from wide-area network cost. In the first, the client runs on the VPS itself, shrinking the network to loopback and isolating server-side and local computation; in the second, it runs on the remote Karachi workstation over the public internet, a realistic deployment with transport included. Both use the liboqs backend, discard a warm-up, and report medians.

One design property shapes everything in these tables: the file never travels to the key service. Only the small wrapped key crosses the network, once, while the file is encrypted locally under the recovered symmetric key. Server-side key-encapsulation cost and network cost are therefore fixed, independent of file size, and only local symmetric encryption grows with the data. Table VIII-a [55] shows the localhost configuration, the tiny fixed key-encapsulation cost set against the linearly scaling symmetric cost.

**TABLE VIII-a. END-TO-END LATENCY, LOCALHOST [55]**

| File size | Server KEM (ms) | AES-256-GCM (ms) | Network (ms) | End-to-end (ms) |
|---|---|---|---|---|
| 1 MB | 0.57 | 6.1 | 32.6 | 42.2 |
| 10 MB | 0.45 | 27.4 | 31.6 | 57.8 |
| 100 MB | 0.38 | 192.7 | 32.7 | 227.8 |
| 500 MB | 0.37 | 998.4 | 34.6 | 1033.3 |
| 1 GB | 0.45 | 2054.3 | 33.0 | 2087.0 |

*Medians, warm-up discarded, liboqs backend. Client co-located on the VPS; the network column reflects loopback through the reverse proxy.*

Key encapsulation remains below one millisecond across three orders of magnitude in file size. At the same time, symmetric encryption grows linearly, exactly as expected when the post-quantum operation protects only the key and never the bulk data. Table VIII-b [56] repeats the measurement from the remote client, separately for encryption and decryption, to include transport cost. The end-to-end figures isolate key establishment and bulk encryption. Client-side signature verification, characterised separately in Table VIII-f [58] at 0.107 ms median, is included in Fig. 6 as a fixed per-object cost.

**TABLE VIII-b. END-TO-END LATENCY, REMOTE (KARACHI TO VPS) [56]**

**Encrypt**

| File size | Server KEM (ms) | AES (ms) | Network (ms) | End-to-end (ms) |
|---|---|---|---|---|
| 1 MB | 0.58 | 3.6 | 564 | 568 |
| 10 MB | 0.61 | 14.5 | 495 | 510 |
| 100 MB | 0.48 | 105.5 | 577 | 695 |
| 500 MB | 0.53 | 542.5 | 474 | 1017 |
| 1 GB | 0.64 | 1147.0 | 471 | 1619 |

**Decrypt**

| File size | Server KEM (ms) | AES (ms) | Network (ms) | End-to-end (ms) |
|---|---|---|---|---|
| 1 MB | 0.60 | 2.6 | 584 | 587 |
| 10 MB | 0.66 | 12.4 | 580 | 592 |

| File size | Server KEM (ms) | AES (ms) | Network (ms) | End-to-end (ms) |
|---|---|---|---|---|
| 100 MB | 0.64 | 108.3 | 618 | 727 |
| 500 MB | 0.56 | 537.0 | 653 | 1196 |
| 1 GB | 0.66 | 1194.8 | 639 | 1835 |

*Medians, warm-up discarded, liboqs backend. Round-trip latency approximately 175 to 185 milliseconds; the network column includes the per-invocation TLS handshake.*

Read together, and as plotted in Fig. 6, the two tables tell the deployment story plainly. For small files, the wide-area network owns the end-to-end time; the SME user is paying for the round trip and the transport handshake, not for cryptography. Only at large sizes does local symmetric encryption take the lead again. In neither regime is the post-quantum key encapsulation a material cost, and that is the central empirical finding here: the overhead of quantum resistance in QCG is negligible against costs an SME already accepts in any cloud workflow.

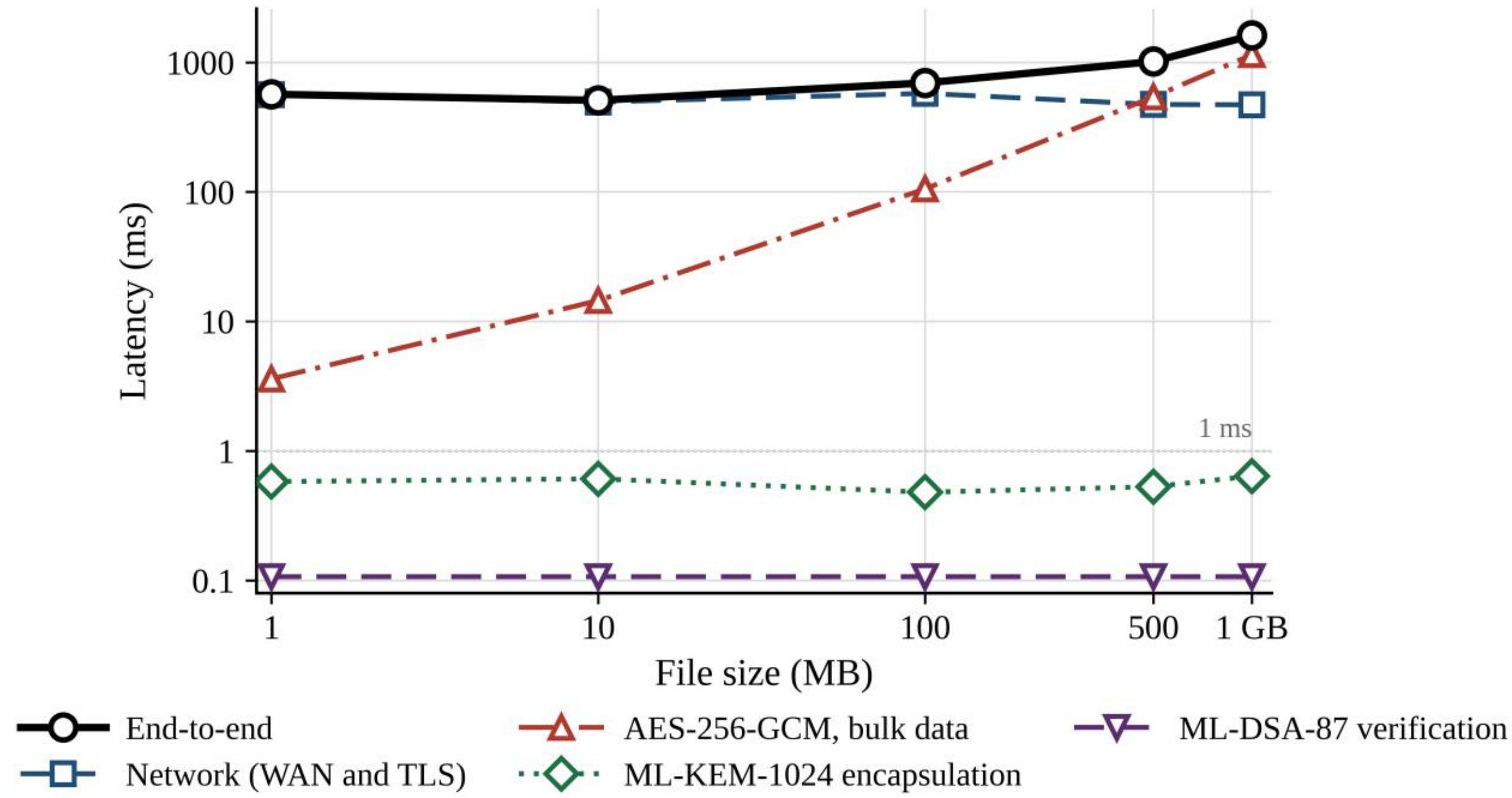

***Fig. 6.*** *End-to-end latency components against file size for the remote configuration. Both post-quantum operations, server-side ML-KEM-1024 encapsulation and client-side ML-DSA-87 verification, stay flat and below one millisecond across all file sizes, while only the bulk symmetric encryption scales with the data and the wide-area network cost remains constant.*

### D. Comparison with a Classical Baseline

To price the transition precisely, ML-KEM-1024 was set against the classical elliptic-curve scheme X25519, in isolation from the network, over one thousand iterations with a warm-up discarded, on the liboqs backend. Both the timing of every operation and the size of every artefact are reported, because a fair comparison must include both. Table VIII-c [57] lists all of it, so no scope is hidden.

**TABLE VIII-c. ML-KEM-1024 VERSUS X25519 [57]**

| Metric | ML-KEM-1024 | X25519 |
|---|---|---|
| KeyGen (ms) | 0.051 | 0.064 |
| Encapsulate (ms) | 0.051 | 0.133 (incl. ephemeral keygen) |
| Decapsulate (ms) | 0.058 | 0.068 |
| Public key (bytes) | 1568 | 32 |
| Ciphertext (bytes) | 1568 | 32 |
| Secret key (bytes) | 3168 | 32 |

*One thousand iterations, warm-up discarded, liboqs backend. The X25519 encapsulation figure includes the ephemeral key generation that an X25519 key-encapsulation construction performs on every message, whereas ML-KEM amortises recipient key generation across many encapsulations; all individual operations are listed so that either a strict per-operation or an as-deployed comparison can be made. Values are medians over one thousand iterations on a shared virtual private server and vary modestly with system load.*

The finding favours the post-quantum scheme, and it has to be stated carefully. The comparison only makes sense at the right granularity. The X25519 encapsulation figure is the cost of a complete key-encapsulation step; the ephemeral key generation that an X25519-based construction performs on every message plus the scalar multiplication; ML-KEM amortises recipient key generation across many encapsulations, so its figure is the per-message cost alone. Measured at that per-message granularity, which matters for systems that establish a fresh wrapped key for every file, ML-KEM-1024 encapsulation was comparable to the X25519 construction, the difference falling within measurement variation. This is not a claim that lattice operations are intrinsically cheaper than elliptic-curve operations; at the level of a single scalar multiplication, they are not. It is the narrower, accurate observation that for the per-message key establishment QCG actually performs, the post-quantum primitive costs nothing extra in computation. The real price of the transition is size. ML-KEM-1024 public keys and ciphertexts run to about 1.5 kilobytes against 32 bytes for X25519, roughly a factor of fifty. This paper does not pretend post-quantum protection is free. The overhead is real; it lives in key and ciphertext size rather than computation time. In QCG's specific setting, the architecture absorbs most of it because exactly one wrapped key of that size crosses the network per operation, and the previous subsection showed network cost at that scale to be bounded by latency, not by bytes.

### E. Symmetric Throughput and Client Hardware Independence

Two further measurements back the claim that QCG runs acceptably on hardware an SME already owns. The first isolates symmetric bulk-encryption throughput on the server, the one term that scales with file size. The second repeats client-side encryption on three genuinely different machines, to show the cost is not tied to expensive equipment. Table VIII-d [55] gives the server throughput.

**TABLE VIII-d. AES-256-GCM THROUGHPUT [55]**

| File size | AES encrypt (ms) | AES decrypt (ms) | Throughput (MB/s) |
|---|---|---|---|
| 1 MB | 6.1 | 4.6 | ~165 |
| 10 MB | 27.4 | 27.4 | ~365 |
| 100 MB | 192.7 | 192.6 | ~520 |
| 500 MB | 998.4 | 998.4 | ~501 |
| 1 GB | 2054.3 | 2054.3 | ~499 |

*Throughput settles at approximately 500 MB/s at scale. Lower throughput on small files reflects fixed per-call overhead that is not amortised over a small quantity of data.*

The client comparison then ran across three machines spanning six years and both major processor vendors: a 2018 ultraportable, a 2020 desktop, and a 2024 high-end laptop, with the server-side key-encapsulation cost near half a millisecond throughout, since it executes on the server rather than the client. Table VIII-e [56] reports client-side symmetric encryption time at several sizes.

**TABLE VIII-e. CLIENT-SIDE ENCRYPTION ACROSS HARDWARE [56]**

| Machine | Class | 10 MB | 100 MB | 500 MB | 1 GB |
|---|---|---|---|---|---|
| i9-14900HX (24c/32t, 31.6 GB DDR5) | 2024 laptop | 14.6 | 105.5 | 542.5 | 1147.0 |
| Ryzen 5 5600X (6c/12t, 15.9 GB DDR4) | 2020 desktop | 10.7 | 191.7 | 459.8 | 1041.2 |
| i5-8365U (4c/8t, 7.8 GB DDR4) | 2018 laptop | 21.6 | 226.3 | 991.5 | (omitted) |

*Medians. The 1 GB point for the i5-8365U is omitted because with 8 GB of memory the machine began paging to disk, which measures storage rather than cryptography; 500 MB is the largest size that runs cleanly on all three machines.*

The comparison produces a finding worth pausing on. The 2020 Ryzen 5 5600X is the fastest of the three at symmetric encryption, roughly 1090 MB/s, edging out the far more expensive 2024 i9 at roughly 950 MB/s. AES-256-GCM here runs single-threaded, so per-core speed and AES-NI matter more than the core count the premium part carries. The oldest machine, the 2018 i5, is about 2.5 times slower at roughly 442 MB/s and still encrypts 100 MB in under a quarter of a second. Across vendors, generations, and price points, there is no hardware barrier on the client side, and the cheapest machine is in fact the quickest at the cryptographic work.

The authenticity layer described in Section VIII signs each public key the key service returns, so its cost is characterised on the same deployment: ML-DSA-87 over one thousand iterations, native liboqs backend, warm-up discarded. Table VIII-f [58] reports the result. Every operation completes well under a millisecond, so the signature adds no meaningful cost to the per-object workflow; as with key encapsulation, the real cost lies in key and signature sizes rather than computation, and those sizes are given beneath the table.

**TABLE VIII-f. ML-DSA-87 SIGNATURE OPERATION COST [58]**

| Operation | Mean (ms) | Median (ms) |
|---|---|---|
| Key generation | 0.109 | 0.104 |
| Sign | 0.244 | 0.210 |
| Verify | 0.118 | 0.107 |

*Medians over one thousand iterations on the shared virtual private server, native liboqs backend, warm-up discarded; values vary modestly with system load. Public key 2,592 bytes, secret key 4,896 bytes, signature 4,627 bytes, the standard FIPS 204 ML-DSA-87 sizes. Signing is marginally slower than verification, as expected for a scheme whose signing step performs rejection sampling while verification is deterministic.*

### F. Abuse Resistance of the Key Service

Sentinel Gate is an abuse-prevention gateway rather than a DDoS mitigation, and the evaluation measures the controls it provides: per-identity rate limiting, resistance to credential stuffing and brute force, and containment of a sustained single-source flood. Because the self-hosted key service is a single point of failure the managed alternatives lack, it was placed under sustained abusive load, with telemetry monitored through the dashboard in Fig. 7, to measure two properties: the proportion of malicious traffic removed before it reaches the key service, and the latency added for a legitimate user during an attack. Since such a gateway separates attacker from user by identity, the attack and the legitimate probe must originate from genuinely different network identities. The attack was therefore generated from the server's own network. At the same time, a paced legitimate probe ran from the Karachi workstation, a distinct public address on a different operator, recording the latency and success of each request over a window covering the attack. Table IX [59] reports the outcome with the gateway disabled and enabled.

**TABLE IX. KEY-SERVICE BEHAVIOUR UNDER LOAD [59]**

| Metric | Gateway OFF | Gateway ON |
|---|---|---|
| Attacker requests sent | 20,077 | 19,005 |
| Attacker requests reaching the KMS | 20,077 (100%) | 219 (1.2%) |
| Malicious traffic blocked | 0% | 98.8% |
| Legitimate success rate (separate IP) | 100% (33/33) | 100% (32/32) |
| Legitimate median latency | 621 ms | 625 ms |
| Legitimate p99 latency | 930 ms | 954 ms |

*Gateway state toggled between runs and verified against the live service configuration. Attack and legitimate probe originate from different public addresses on different networks.*

Two results are clean and repeatable. The gateway removed 98.8% of the attack traffic before it reached the key service, against nothing with the gateway off, and it held on every run. And it cost the legitimate user nothing perceptible: a 625 millisecond median with the gateway on, against 621 without, is within noise.

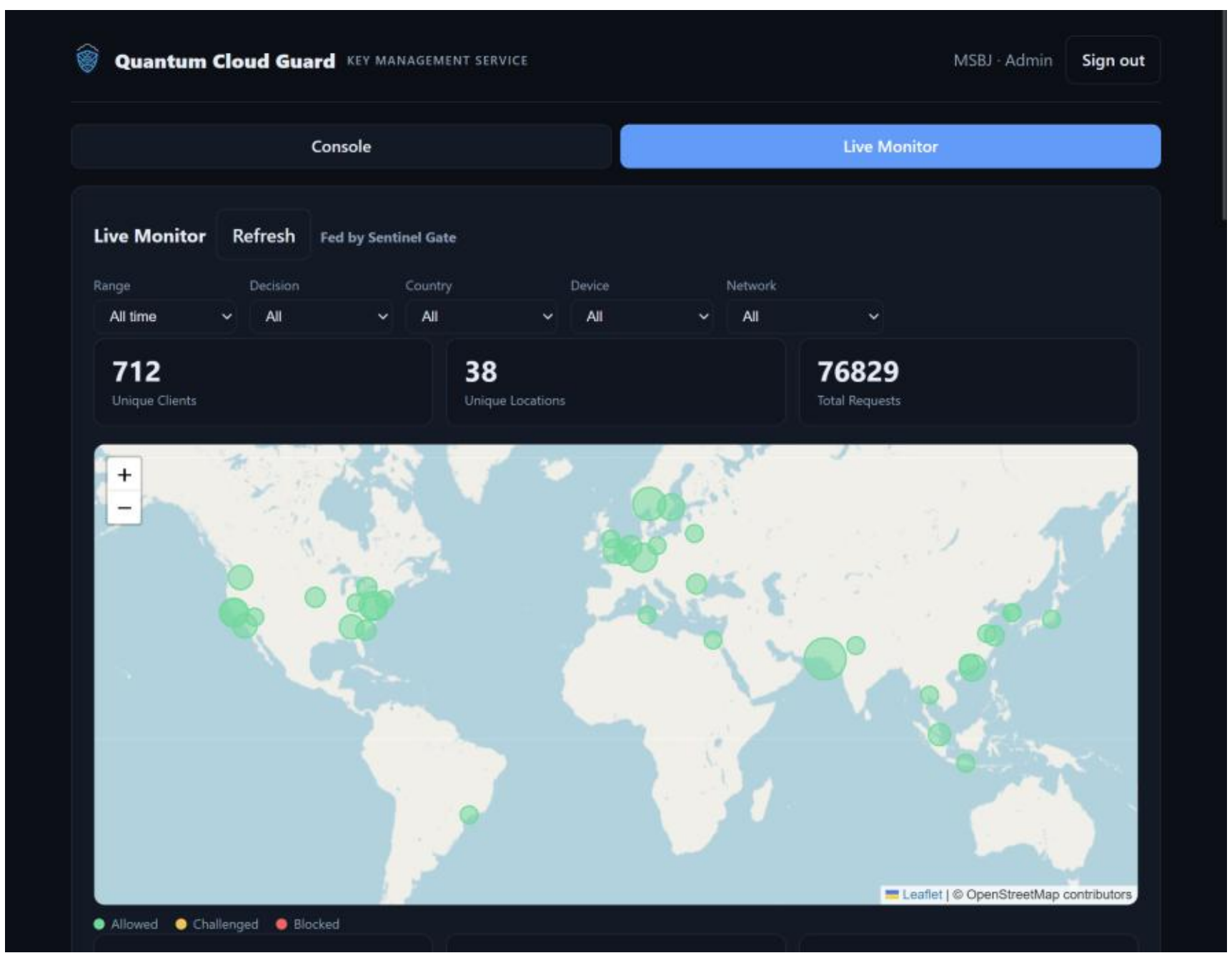

***Fig. 7.** Live monitoring dashboard of the key service, fed by the Sentinel Gate gateway, showing per-request geolocation, device and network classification, and country distribution used for abuse detection.*

The scope of this result is stated honestly. The legitimate user succeeded on every request in both configurations, so the experiment did not demonstrate prevention of an availability collapse; none occurred, because a single attacking machine could not generate enough load to exhaust the server. Raising concurrency from sixty workers to two hundred moved the achieved rate no higher than roughly 440 to 640 requests per second, a ceiling set by the single path and single source address of one machine, not by the server. A single-source flood of this size is precisely what the layered rate limiting is built to neutralise, and it did. Demonstrating a prevented collapse would need genuinely distributed load from many addresses, beyond this evaluation's resources. The paper therefore claims what the data support: near-total filtering of a sustained single-source flood at negligible latency cost, and does not claim distributed denial-of-service resilience, which sits outside Sentinel Gate's design and would be layered beneath it at the network level; extending these controls to distributed abuse is future work.

The failure to force a collapse from one machine is itself evidence for the design. The layered defence neutralises a single-source flood so completely that one source cannot push enough traffic to harm.

#### G. Economic Feasibility

The last axis is cost, since affordability is a central claim. The comparison sets the self-hosted model, a key service on a fixed-price virtual private server, against the metered managed key services of the two largest cloud providers. Provider figures come from the published pricing pages current as of July 2026 and are reproduced in Table X; the self-hosted figure is the actual monthly cost of the Hetzner CX22 this paper's deployment runs on.

**TABLE X. MONTHLY COST COMPARISON VERSUS MANAGED KMS**

| Cost factor | QCG (self-hosted) | AWS KMS | Google Cloud KMS |
|---|---|---|---|
| Infrastructure base | EUR 4.49 flat (VPS) | none | none |
| Per-key or per-version | included | USD 1.00 per key | USD 0.06 per key version |
| Key rotation | free, unlimited | USD 1.00 per key, first two rotations | free (re-encryption billed as operations) |
| Free operation allowance | not applicable (unmetered) | 20,000 requests | 10,000 operations |
| Per-operation charge | EUR 0 (unmetered) | USD 0.03 per 10,000 | USD 0.03 per 10,000 |
| Asymmetric operation charge | EUR 0 (unmetered) | USD 0.15 per 10,000 | USD 0.15 per 10,000 (HSM) |
| Illustrative: 10 keys, 1M operations/month | EUR 4.49 flat | approx. USD 42 | approx. USD 30 |
| Key custody separated from data host | Yes | No | No |
| Cost model | Flat and fixed | Metered and variable | Metered and variable |

*Provider figures from the AWS Key Management Service and Google Cloud Key Management Service pricing pages, current at the time of writing. AWS charges USD 1 per key per month and adds USD 1 per month for each of the first two rotations of a key, capped thereafter, with 20,000 free requests per month and USD 0.03 per 10,000 requests above that. Google Cloud charges per active key version, USD 0.06 per month in its own worked example, with 10,000 free cryptographic operations per month and USD 0.03 per 10,000 thereafter. The Hetzner CX22 instance is a flat EUR 4.49 per month and includes 20 TB of traffic and network-level protection at no additional charge. Illustrative totals are order-of-magnitude figures for a small deployment and will vary with exact usage; the AWS figure assumes ten keys with two rotations and roughly one million symmetric operations after the free tier, and the Google figure assumes ten keys with several active versions and a comparable operation count.*

Three points follow, and none of them is that the managed services are expensive; at low request volumes they are not. The first is the structure of the cost rather than its magnitude. The managed model is metered, scaling with keys, key versions, and operations. Hence, a workload that rotates keys aggressively or runs many operations pays for its own hygiene, while the flat self-hosted model absorbs the same hygiene for nothing extra. The second is trajectory. A metered cost recurs, grows with usage, and is set by a single provider with pricing power over a service the customer cannot easily leave; the flat cost of a self-hosted instance is exposed only to the competitive general hosting market. The third reduces to price at all. Under the managed model, the keys live with the same provider that holds the encrypted data, which recreates precisely the trust concentration QCG exists to break, so the managed service cannot offer the custody separation this architecture is founded on at any price. The economic case is therefore not that QCG is cheaper in every instance. It is that the cost is predictable, good key hygiene is never penalised, and key custody stays with the enterprise, which the metered alternatives structurally cannot offer.

#### H. Summary of Findings

The evaluation carries the three contributions on measured evidence. The post-quantum primitive is cheap even on a microcontroller, so no hardware bars an SME client. The end-to-end overhead is a small fixed quantity buried under network and bulk-encryption costs an SME already accepts, and key establishment matches the classical baseline, its only real cost being key and ciphertext size, which the architecture largely absorbs. The self-hosted key service, single point of failure though it is, filtered almost all of a sustained single-source flood at no latency cost to legitimate users, with distributed-attack resilience left to future work. The economic side, the retained custody and the flat cost, is taken up in the discussion that follows.

## X. Implementation Challenges and Deployment Considerations

Section VIII shows that QCG can be built from free components on commodity infrastructure. A proof of concept is not a production system, though, and honesty about the distance between the two belongs in any evaluation worth trusting. This section sets out the main engineering challenges met while building QCG, the decisions taken in response, and the hardening an SME production deployment would still need. The point is to mark precisely what the present work demonstrates and what it does not.

### A. The Backend Portability and Performance Trade-off

The most consequential implementation decision was the post-quantum backend. Two options existed: a pure-Python ML-KEM, which installs with no system dependencies and runs wherever a Python interpreter does, and native liboqs, written in C, substantially faster, and needing compilation or a platform-specific binary. The tension maps straight onto the SME setting. No in-house expertise argues for zero dependencies; performance argues for native code. QCG refuses to choose for the operator: both backends live behind one interface, the pure-Python one is the default for portability, and liboqs is a switch away where performance matters. The measurements in Section IX use liboqs, because that is what an SME would deploy once past its first trials, and because the per-file key operation is dwarfed by network cost anyway, so the choice barely moves the end-to-end numbers. The lesson travels beyond this system. Post-quantum migration for resource-limited organisations is better served by implementations that degrade gracefully to a dependency-free mode, even at a performance cost which, in a network-bound workload, is largely invisible.

### B. State, Sessions, and the Cost of Statelessness

A key service that authenticates users and holds decrypted key material in memory has to manage session state, and doing that correctly across restarts and concurrent requests is a classic source of quiet error. QCG pushes session and gateway state out to Redis instead of holding it in process memory, which keeps the key service itself nearly stateless and lets the rate-limit counters and the token store survive an application restart. The price is one more component to deploy and secure, and a dependency whose availability now matters. For an SME, the trade is acceptable: Redis is mature and cheap to operate, but it is a real addition to the trust and maintenance surface that a single-binary design would avoid. It is recorded here rather than smoothed over.

### C. The Master Passphrase as a Single Point of Human Failure

The vault sealing the secret key answers to a master passphrase run through Argon2id. Cryptographically, that is sound; practically, it relocates a large share of the system's security into a secret a human chooses. A weak passphrase undercuts the memory-hard protection, and a lost one makes the sealed vault unrecoverable forever, because the passphrase is deliberately stored nowhere. QCG blunts the first risk with Argon2id's cost parameters, which keep brute force expensive even against a stolen vault, and the second with the property that the encrypted vault can be backed up freely without exposing the key. Hence, recovery from losing the infrastructure stays possible while recovery from losing the passphrase does not. A production deployment would add organizational controls on top:

passphrase policy, secure escrow of recovery material, but those are operational practice, not cryptographic design, and they sit outside this paper.

### D. What Production Hardening Would Still Require

Several properties a production key service is expected to have are deliberately out of scope, and they are listed plainly so the evaluation is not mistaken for a claim of production readiness. High availability is absent: there is one key service instance, and losing the VPS halts key operations until recovery from backup. Audit logging exists, but formal compliance certification does not. And the abuse-prevention gateway, as Section IX states, is no defence against distributed denial-of-service at network scale, which an SME would buy from its hosting provider or a network-level service. None of these gaps touches the central claims, which concern the feasibility, cost, and abuse resistance of the architecture. Each would need to be closed before an SME trusted QCG with production data, and stating this is part of accurately describing the contribution.

## XI. Security Analysis

This section makes the security argument explicit, and it starts where any such argument must: with the adversary. It states the threat model, argues confidentiality by reduction to standardised components, treats the server as a decapsulation oracle, and then addresses the man-in-the-middle, downgrade, key-compromise, and session-token threats in turn. The analysis is semi-formal by intent. It does not re-prove ML-KEM-1024 or AES-256-GCM, which are standardised and proven elsewhere; it argues that their composition in QCG preserves the properties the architecture depends on.

### A. Threat Model

The adversary is a computationally bounded attacker who will, within the lifetime of the protected data, gain access to a cryptographically relevant quantum computer: the harvest-now-decrypt-later adversary, recording ciphertext today to decrypt it once the capability matures. Four capabilities are assumed:

**(1) Cloud access.** complete and permanent access to everything held in the public cloud, treated throughout as an untrusted host of ciphertext.

**(2) Network control.** the full powers of an active network attacker: observing, recording, intercepting, and tampering with client-to-service traffic.

**(3) Interface access.** the ability to interact with the exposed key-service interface as any network party can.

**(4) A quantum computer.** by definition, arriving within the data's confidentiality lifetime.

The boundaries are stated as plainly as the capabilities. The key service runs on infrastructure under the SME's exclusive administrative control, and that infrastructure sits inside the trust boundary: an adversary who takes administrative control of the VPS, and with it the secret key in volatile memory during an active session, is outside the threat model, in exactly the way physical compromise of an organisation's own servers is outside the threat model of most systems. That is the precise sense in which QCG trusts the key infrastructure and never the storage layer. The user's endpoint is trusted on the same footing; an adversary already on the machine that runs the client can read plaintext directly, and no cloud encryption scheme can protect data that is exposed on the very device encrypting it. These boundaries, summarised in Table XI and drawn in Fig. 8, are not evasions. They are the standard delineation for a system of this class: QCG defends the confidentiality of data at rest in an untrusted cloud against a future-quantum network adversary, and claims nothing about the compromise of the SME's own trusted infrastructure or endpoints.

**TABLE XI. THREAT MODEL SUMMARY: ADVERSARY CAPABILITIES AND TRUST BOUNDARIES**

| **Element** | **Position in the threat model** |
|---|---|
| Adversary | Computationally bounded attacker who gains a cryptographically relevant quantum computer within the data lifetime (harvest-now-decrypt-later) |
| Capability 1 | Complete and permanent access to all ciphertext stored in the public cloud |
| Capability 2 | Active network attacker: observes, records, intercepts, and tampers with client-to-service traffic |
| Capability 3 | Interacts with the exposed key-service interface as any network party |
| Inside trust boundary | The self-hosted key service and the secret key in its volatile memory; the user endpoint holding plaintext and passphrase |
| Outside scope | Administrative compromise of the SME's own server; compromise of the user's endpoint device |
| Defended property | Confidentiality of data at rest in an untrusted cloud against a future-quantum network adversary |
| Authenticity mechanism | Client-verifiable ML-DSA-87 signatures on every served public key, checked by the client before any encapsulation |

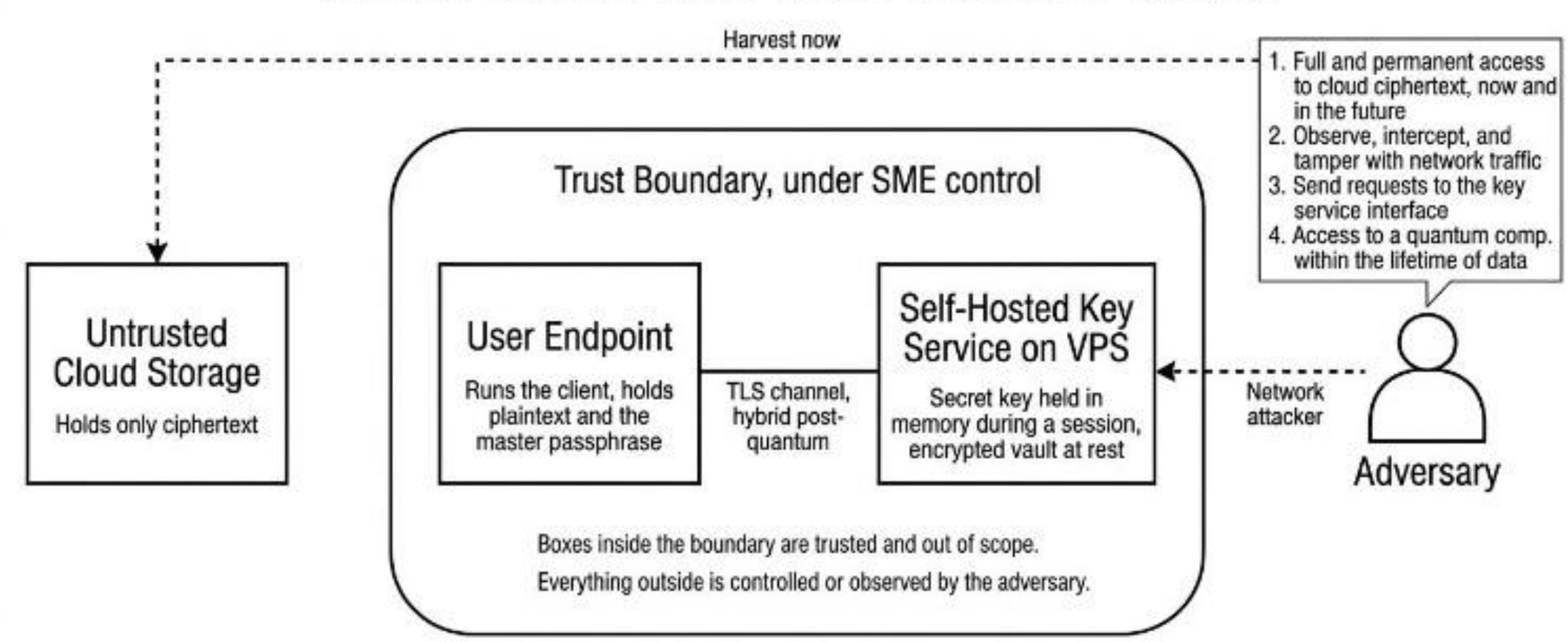

***Fig. 8.*** *System security and trust boundary model. The user endpoint and the self-hosted key service lie inside the trust boundary under SME control, while the cloud storage, the network, and the adversary lie outside it.*

**B. Confidentiality of the Core Construction**

**Claim 1 (Confidentiality).** *Under the threat model of Subsection A, and assuming the IND-CCA2 security of ML-KEM-1024 and the authenticated-encryption security of AES-256-GCM, no computationally bounded adversary, classical or quantum, recovers any plaintext from the stored ciphertext tuple (c_kem, N, c_dem, T) without the recipient secret key.* The argument is a reduction: any adversary who breaks this confidentiality breaks one of the two standardised primitives it is built on.

The confidentiality claim stands on the KEM-DEM construction of Layer 1. A fresh data key encrypts each object with AES-256-GCM; that data key is encapsulated under the recipient's ML-KEM-1024 public key; and the security of the whole is the security of the standard, well-understood hybrid KEM-DEM paradigm. AES-256 in Galois/Counter Mode provides authenticated encryption, confidentiality, and integrity together under its key, and the 256-bit key provides a comfortable margin against Grover's algorithm [19], which weakens a symmetric key by at most a square-root factor and so leaves AES-256 at roughly 128 effective post-quantum bits. The data key itself sits behind ML-KEM-1024, standardised in FIPS 203 and IND-CCA2 secure [48] under the hardness of module learning with errors, a problem with no known efficient algorithm, classical or quantum. A uniformly random key encapsulated under an IND-CCA2 mechanism gives the holder of the ciphertext tuple nothing beyond what a random guess would, and therefore nothing about the plaintext under it. The composition inherits the confidentiality of AES-256-GCM under a key the adversary cannot recover, and the harvest-now-decrypt-later adversary loses on both layers at once: the symmetric outer layer and the key-encapsulation inner layer are each quantum-resistant, so ciphertext recorded today decrypts to nothing in a post-quantum future.

### C. The Key Service as a Decapsulation Oracle

One feature of QCG asks for its own analysis. Once a session exists, the key service decapsulates submitted ciphertexts and returns the shared secret, a decapsulation oracle in the precise cryptographic sense. Can an adversary holding that oracle extract the secret key, or learn about other encapsulations? The answer falls straight out of IND-CCA2. That security notion is exactly the guarantee that an adversary who submits chosen ciphertexts to a decapsulation oracle and studies the results cannot distinguish the shared secret of a target ciphertext from random, and cannot recover the secret key. What the QCG service returns for a submitted ciphertext is the shared secret of that ciphertext and nothing else; the ciphertext itself is already public once transmitted, so the information corresponds only to what the adversary already holds. The residual risk is therefore not cryptographic, since IND-CCA2 forecloses key extraction. It is the access-control risk of an adversary opening a session in the first place and misusing legitimate decapsulation, which the session-token protections below govern, and which presupposes an uncompromised endpoint throughout.

### D. Man-in-the-Middle and Downgrade Resistance

Authenticity of the recipient public key is what stops a man-in-the-middle substituting its own key and letting the client encapsulate under it. QCG binds public keys with an ML-DSA-87 signature; the client verifies it before encapsulating, and an adversary who cannot forge the signature cannot substitute the key. Because the scheme is post-quantum, this authenticity survives the same quantum adversary the confidentiality design anticipates. Downgrade resistance is structural rather than watched-for. As Section VII describes, QCG carries no classical key-establishment algorithm, no negotiation step, and no legacy fallback: the sole mechanism is ML-KEM-1024 encapsulation, the sole cipher AES-256-GCM, and any object of the exact expected structure is rejected. With no weaker mode inside the suite, there is nothing to force a downgrade towards, which closes the path negotiation-based hybrids leave open. The price of removing negotiation is algorithm agility, and in the SME setting that trade is right: one fixed, known-good suite is simpler and harder to misconfigure than a negotiable one.

### E. Key Compromise and Revocation

Two kinds of key compromise need separating. A per-object data key, if it ever leaked, exposes exactly one object, because the envelope design draws a fresh key per object and never reuses one; the blast radius is one file. The recipient secret key is the serious case, since it unwraps every data key. QCG keeps it sealed in the encrypted vault at rest, so compromising it means either defeating the Argon2id-protected vault, which the cost parameters make impractical, or lifting it from the volatile memory of the key service mid-session, which requires administrative compromise of the SME's own trusted infrastructure and sits outside the threat model. When a secret key has to go, on suspicion of compromise, a personnel change, or policy, QCG rotates the recipient keypair: new objects wrap under the new key, existing objects keep the key they were sealed with, retained for their decryption. The per-object envelope already confines any single data-key exposure, and rotation caps how long any one secret key protects new material. Hence, the design gives fine-grained control over exposure without re-encrypting a byte of stored data.

### F. Session Token Security

After authentication, everything runs on the session token, and the practical security of an established session lives in how that token is protected. In QCG, it is a high-entropy value from a cryptographically secure generator, held server-side with a bounded lifetime, carried only over TLS: not guessable, not valid forever. Three threats matter. Guessing fails on entropy alone. Replay is bounded by the limited lifetime and the transport protection of the channel, and can be tightened further by binding the session to its originating context. Interception is defeated by the confidentiality of the TLS channel itself, and this is where hybrid post-quantum TLS stops being a detail: a channel carrying authentication and token traffic under a purely classical key exchange would hand that traffic to the harvest-now-decrypt-later adversary, reintroducing at the transport layer the exact threat the application layer exists to defeat. The live deployment closes this directly. Its TLS terminator negotiates the X25519MLKEM768 hybrid group, the construction major providers now deploy [49], combining classical X25519 with the FIPS 203 mechanism ML-KEM-768. Hence, the channel holds as long as either component holds, and a verification against the running service confirms that a supporting client negotiates this group rather than a classical one. The transport protection comes from the TLS stack, distinct from and independent of the application-layer protection that is this paper's contribution, and together they leave the system post-quantum at both layers. The session's security therefore rests on a token bounded in time and protected in transit, both of which the design provides, and on an uncompromised endpoint, which the threat model assumes.

One hardening is named rather than claimed: the implementation does not yet bind a token to a client network address or device fingerprint. Binding would raise the cost of replaying a stolen token from another origin, and it belongs on the production-hardening list rather than among the properties established here.

### G. Summary of the Security Argument

Under the stated threat model, the security of QCG reduces to the security of its standardised components and the correctness of their composition. Confidentiality against the future-quantum network adversary follows from the IND-CCA2 security of ML-KEM-1024 and the authenticated encryption of AES-256-GCM, composed in the standard KEM-DEM paradigm. The same IND-CCA2 property disarms the decapsulation oracle. Authenticity rides on post-quantum signatures over public keys, and downgrade dies structurally for want of a weaker mode to force. What remains is operational, not cryptographic: tokens, which the design bounds in time and in transit, and the integrity of the SME's own infrastructure and endpoints, which the model places inside the trust boundary. QCG claims nothing against an adversary who controls that infrastructure or those endpoints; within its stated boundaries, its confidentiality rests on assumptions no weaker than those of the primitives it composes.

## XII. Discussion and Future Work

The mechanics settled, three questions decide whether QCG matters in practice: how it stands against the managed post-quantum key services that have just arrived, what systemic risk an all-lattice design carries, and where the work goes next, with a return to the longer horizon of the economic argument from Section IX.

### A. Self-Hosted Custody Against Managed Post-Quantum Key Services

The most immediate challenge to the premise is that the major cloud providers now sell post-quantum key management as a managed service [49], [50], which invites an obvious question: why would an SME self-host? Because of the trust model, and the answer does not reduce to price. A managed key service puts the keys under the administrative control of the same provider that hosts the encrypted data, reuniting in one trust domain exactly the two things QCG separates. For harvest-now-decrypt-later, that separation is the whole point. If the provider storing the ciphertext also holds the keys, then it, or anyone who compels or compromises it, holds both halves, and the client-side encryption has protected nothing the provider cannot undo itself. Self-hosting on infrastructure the SME controls is what keeps custody of the keys apart from custody of the data. The managed services are convenient, and at low volume they are cheap. But convenience bought by recombining keys and data is a trade the QCG threat model cannot accept, which is why self-hosted custody, not any primitive, is the sharpest contribution of this architecture.

### B. The Systemic Risk of an All-Lattice Design

One limitation deserves candour: QCG draws both confidentiality and authenticity from lattice-based primitives, ML-KEM-1024 for key establishment and ML-DSA-87 for authenticity, both resting on structured lattice problems. A fundamental advance against module learning with errors would weaken both at once, where a design with independent assumptions would lose only one. QCG accepts this with open eyes. Lattice problems are among the most heavily studied in cryptography, standardised precisely because they survived that scrutiny, and the AES-256 outer layer remains an independent symmetric line of confidentiality that a lattice break would not by itself defeat. For authenticity, the hedge is concrete: pairing the lattice signature with a hash-based scheme such as SLH-DSA, whose security rests on nothing more than hash-function security, would remove the single point of dependence, and it is the natural move wherever diversity of assumptions is demanded. The same hedge for confidentiality is harder, because no comparably efficient key-establishment mechanism yet rests on assumptions independent of both lattices and classical number theory. That is an open problem for the field, not for QCG in particular.

### C. Key Rotation in the Envelope Model

The envelope design leaves QCG with an unusually light rotation burden, worth saying because rotation is usually a costly obligation. Every object is sealed under its own fresh data key, so the data keys never rotate at all: no long-lived key sits over a large volume of data waiting to be replaced, because exposure is already confined to one object per key. Rotation applies only to the recipient keypair that wraps the data keys, and it serves compromise response, personnel change, and compliance rather than routine blast-radius control. That is lighter and more predictable than designs encrypting data directly under a few long-lived keys, and it re-encrypts nothing, since rotation governs only how new objects are wrapped. In contrast, existing ones keep the keys they were sealed with. For an SME, aggressive key hygiene costs no performance and no re-encryption, the exact opposite of a metered service that bills every key and every rotation.

### D. The Longer-Term Economic Argument

Section IX showed QCG to be cost-predictable rather than merely cheap, and the distinction grows sharper over a longer horizon. A managed service is a metered recurring cost, scaling with keys, versions, and operations, set by one provider with pricing power over a service that is hard to leave. A self-hosted service is a flat cost exposed only to a competitive hosting market, where switching, while not trivial, is far cheaper than exiting a proprietary key service. Over the multi-year lifetime of exactly the long-lived data that motivates post-quantum protection, a metered cost compounding under one provider and a flat cost sitting in a competitive market diverge materially, whatever the two look like at any single moment. Add that the flat model never penalises the frequent rotation good hygiene recommends, and the economic case lands where it did in Section IX: not that self-hosting always wins on price, but that it is predictable, aligns cost with good practice, and keeps the custody separation the security model stands on.

Table II compared QCG with prior research; Table XII completes the picture with the practical systems an SME would actually weigh: the managed cloud key services whose pricing Section IX examined, and the most widely used self-hosted secrets manager. The table assesses these against the properties QCG is built to satisfy. Entries reflect the state of the systems at the time of writing, and the post-quantum features of the self-hosted alternative in particular are experimental and signature-oriented, not the client-side hybrid data encapsulation QCG provides.

**TABLE XII. COMPARISON OF QCG WITH PRACTICAL KEY-MANAGEMENT SYSTEMS**

| Property targeted by QCG | QCG | Managed cloud KMS (AWS, Google) | Self-hosted secrets manager (HashiCorp Vault) |
|---|---|---|---|
| Client-side hybrid post-quantum data encryption | Yes: ML-KEM-1024 with AES-256-GCM at the client | Post-quantum protection is at the transport layer of the service, not client-side data encapsulation. | AES-256-GCM barrier at rest; post-quantum support is experimental and signature-oriented in the paid edition |
| Key custody separated from data custody | Yes: keys held by the SME, data in a separate untrusted cloud | No: the provider holds the keys and commonly also the data | Yes when self-hosted: the operator holds the keys |
| Software-only, no specialised hardware | Yes: commodity virtual private server | Managed service; hardware abstracted but under provider control | Yes: runs on commodity infrastructure |

| Property targeted by QCG | QCG | Managed cloud KMS (AWS, Google) | Self-hosted secrets manager (HashiCorp Vault) |
|---|---|---|---|
| Cost model | Flat and predictable, independent of key count and operations | Metered: per key, per key version, and per operation | Open-source core is free to self-host; advanced and post-quantum features require the paid edition |
| Open and inspectable implementation | Yes: source publicly available | No: proprietary service | Core is open-source; enterprise features are proprietary |
| Integrated application-layer abuse prevention | Yes: built-in gateway (per-identity limiting, brute-force resistance) | Provider handles infrastructure protection; not an SME-controlled application-layer control | Access control and policies present; abuse-prevention gateway not an integrated equivalent |
| Post-quantum authenticity of served keys | Yes: every public key returned is signed with ML-DSA-87 and verified by the client before use. | Partial: ML-DSA offered as a customer-managed signing primitive; authenticity of served material rests on TLS and provider trust. | Partial in the paid edition: post-quantum signing exists as a primitive; served material is not signed for client verification |

**E. Future Work**

Several directions follow naturally. The gateway wants evaluation against genuinely distributed abuse and integration with the network-level protections beneath its application-layer scope. The all-lattice dependence shrinks with the hash-based signature hedge above. The single key-service instance becomes highly available through replication, closing the availability gap Section X concedes. The evaluation here characterises single-client latency, server throughput, and hardware independence, not the service under many concurrent clients, so a study of throughput and latency under sustained multi-client, enterprise-scale load is the natural extension; it would locate the point where one instance needs horizontal scaling and would shape the replication design. And beyond the architecture, the most significant direction is physical: the embedded benchmark measured what ML-KEM-1024 costs on a microcontroller, not what it leaks there. A leakage-aware analysis of the standardised schemes on constrained hardware, quantifying and localising side-channel exposure and validating countermeasures, is the subject of ongoing work and the natural sequel to this study.

## XIII. Conclusion

This paper has presented QCG, a software-only hybrid post-quantum encryption architecture for the small and medium-sized enterprise, the segment most exposed to harvest-now-decrypt-later and least served by what exists. The contribution is not new cryptography. It is the integration of standardised primitives into a deployable architecture uniting three properties no prior SME-focused system combines: client-side hybrid encryption sealing data under AES-256-GCM and ML-KEM-1024 before it reaches an untrusted cloud; a self-hosted key service that keeps custody with the enterprise rather than the provider storing the data, and signs every public key it serves with ML-DSA-87; and an integrated abuse-prevention gateway in front of that service. The evaluation put measured evidence under each property, on real infrastructure and constrained hardware. The post-quantum primitive costs roughly forty milliseconds on a microcontroller worth a few tens of dollars and matches the classical baseline at the per-message granularity used; its only material cost is the size of keys and ciphertexts. The end-to-end overhead is a small fixed quantity buried under costs an SME already accepts. The gateway removes almost all of a sustained single-source abuse flood at no perceptible cost to legitimate users. The security analysis argued confidentiality by reducing it to the proven security of the standardised components under an explicit threat model, and the economics showed the protection deployable at a flat, predictable cost with the custody separation intact. Together, the results carry the thesis: quantum-safe protection of cloud data is computationally, operationally, and economically feasible at SME scale, and the obstacle to adoption was never the cryptography but the absence of an architecture built around the constraints of the organisations that most need it.

The claim is bounded on purpose. This paper does not resolve the whole post-quantum migration problem for small enterprises, provide key management certified for regulated industries, or address network-layer distributed denial-of-service protection, multi-party or threshold key management, or formal verification of the implementation. What it claims, the evidence supports: a complete, deployable, computationally feasible architecture for client-side post-quantum encryption with self-hosted key custody can be assembled from standardised primitives and commodity infrastructure, and it is viable at the scale of the small enterprise. QCG is offered as one such architecture, its implementation open, as a basis for others to build on.

## Declaration of Competing Interest

The author declares that he has no known competing financial interests or personal relationships that could have appeared to influence the work reported in this paper. This research received no specific grant from any funding agency in the public, commercial, or not-for-profit sectors.

## Data Availability

The complete QCG implementation, the deployment configuration, and the scripts used to produce every benchmark reported in this paper are openly available in the public repository cited in Section VIII, in which the gateway [39] and the key service [46] respectively reside as separate project folders, so that all results can be independently reproduced.